\documentclass{elsart}

\usepackage{graphicx,color}
\usepackage{amsmath}
\usepackage{amsfonts, amssymb, amstext, amsxtra,times}

\newcommand{\be}{\begin{equation}}
\newcommand{\ee}{\end{equation}}
\newcommand{\eqr}[1]{Eq.~(\ref{#1})}

\DeclareMathOperator{\Tr}{tr}

\DeclareMathOperator{\iim}{Im}
\DeclareMathOperator{\rre}{Re}

\newcommand{\al}{\alpha}
\newcommand{\ep}{\epsilon}

\begin{document}

\begin{frontmatter}

\title{Effective Equilibrium Theory of Nonequilibrium Quantum Transport}
\author[Yale]{Prasenjit Dutt}, \author[Northwestern]{Jens Koch}, \author[SUNY]{Jong Han} and \author[Yale]{Karyn Le Hur}

\address[Yale]{Departments of Physics and Applied Physics, Yale University, New Haven, CT 06520, USA}

\address[Northwestern]{Department of Physics and Astronomy, Northwestern University, Evanston, IL  60208, USA}

\address[SUNY]{Department of Physics, State University of New York at Buffalo, Buffalo, NY 14260, USA}

\begin{abstract}

The theoretical description of strongly correlated quantum systems out of equilibrium presents several challenges and a number of open questions persist. In this paper we focus on nonlinear electronic transport through an interacting quantum dot maintained at finite bias using a concept introduced by Hershfield [Phys.\ Rev.\ Lett.\ \textbf{70}, 2134 (1993)] whereby one can express such nonequilibrium quantum impurity models in terms of the system's Lippmann-Schwinger operators. These scattering operators allow one to reformulate the nonequilibrium problem as an effective equilibrium problem associated with a modified Hamiltonian. In this paper we provide a pedagogical analysis of the core concepts of the effective equilibrium theory. First, we demonstrate the equivalence between observables computed using the Schwinger-Keldysh framework and the effective equilibrium approach, and relate the Green's functions in the two theoretical frameworks. Second, we expound some applications of this method in the context of interacting quantum impurity models. We introduce a novel framework to treat effects of interactions perturbatively while capturing the entire dependence on the bias voltage. For the sake of concreteness, we employ the Anderson model as a prototype for this scheme. Working at the particle-hole symmetric point, we investigate the fate of the Abrikosov-Suhl resonance as a function of bias voltage and magnetic field. 

\end{abstract}

\end{frontmatter}

\pagebreak

\section{Introduction}

Quantum systems exhibiting an interplay of interactions and out-of-equilibrium effects are of significant interest and constitute an active area of research
\cite{Goldhaber,Kontos,Egger,Yacoby,LeHur_Frac,Mora_LeHur,Bockrath_2010,McEuen,Mason,Mirlin}. In contrast to the equilibrium setting, the theoretical foundations of these systems are yet to be well understood. Questions such as whether renormalization group and universality concepts can be generalized to the nonequilibrium domain persist. Furthermore, there is a lack of a unifying framework for treating such systems. The real-time Schwinger-Keldysh formalism \cite{Schwinger,Keldysh,Rammer,Meir_Wingreen} has been successfully employed to study certain classes of problems. Most such treatments have been perturbative and is often a delicate procedure. Care must be taken to avoid the appearance of infrared divergences, an artifact of the the infinite contour of integration \cite{Parcollet_Hooley}. While quantum impurity models in equilibrium have been extensively studied \cite{Hewson}, probing and understanding properties of nonequilibrium steady states is a far more subtle task. Nevertheless, there have been significant advances in our understanding via several distinct approaches, which include the scattering Bethe Ansatz \cite{Mehta_Andrei,Mehta_Andrei_new}, field theory techniques \cite{Fendley,Boulat_2008}, time-dependent density matrix renormalization group (RG) \cite{Boulat_2008,Dagotto}, time-dependent numerical RG \cite{Anders_2008,Anders_2010}, perturbative RG \cite{Rosch,Chung,schoeller1}, Hamiltonian flow equations \cite{Kehrein}, functional RG \cite{Schoeller,Karrasch,Wolfle}, strong-coupling expansions \cite{Mora_2009,LeHur_2008,Mora_2008,Mitra}, master equations \cite{Timm_2010}, diagrammatic Monte Carlo \cite{Fabrizio,Millis,Eckstein,Muehlbacher,Millis_review} and imaginary-time nonequilibrium quantum Monte Carlo \cite{Han_2007,Han_2010}.

Mesoscopic quantum systems such as quantum dots (``artificial atoms''), which are characterized by a set of discrete energy levels, when coupled to reservoir leads \cite{Glazman} are described by quantum impurity models. In this context, the Anderson model, which describes a single discrete level coupled via tunneling to a Fermi-liquid sea, is of particular interest \cite{Hewson}. For a level with sufficiently low energy and with strong on-site interaction, the Anderson model mimics a Coulomb-blockaded quantum dot \cite{Devoret}. In this regime, charge degrees of freedom on the level are frozen out and the Anderson model becomes intimately related to the Kondo model. The latter describes a magnetic impurity (manifested by the spin of the highest occupied level for an odd number of electrons on the quantum dot) entangled with the spins of a Fermi sea (In the mesoscopic setting, the reservoir leads play the role of Fermi seas). The Kondo entanglement then produces a prominent Abrikosov-Suhl resonance in the density of states of the quantum dot at the Fermi level, which results in perfect transparency when the quantum dot is coupled  symmetrically to the source and the drain \cite{Glazman,Goldhaber_1998,Delft,Weis,schoeller2}. Nanoscale systems can be routinely driven out of equilibrium by applying a bias voltage between the two reservoir leads. Among the various issues which arise out of equilibrium, the precise fate of the Abrikosov-Suhl resonance \cite{Leturcq} when applying a finite bias voltage, remains an issue of debate. In general, it seems essential to elaborate theoretical and numerical methods which will allow access to the full current-voltage characteristics of interacting mesoscopic systems.

The main purpose of this paper is to reformulate electronic nonequilibrium transport in quantum impurity models in terms of an effective equilibrium steady-state density operator. This concept was initiated by Hershfield \cite{hershfield_reformulation_1993} who proposed that the appropriate density operator can be expressed in terms of the system's Lippmann-Schwinger operators in a Boltzmann form. While noninteracting and effectively noninteracting theories \cite{Hershfield_Schiller} have been treated analytically within this framework, extensions to interacting models have yet to be explicitly demonstrated. Recently, there have been several numerical efforts at implementing this scheme \cite{Anders_2008,Anders_2010,Han_2007,Han_2010}. Also, the interacting resonant level model has been treated perturbatively using the steady-state density matrix along with dynamical ``impurity conditions'' \cite{doyon_new_2007}, effectively circumventing explicit construction of Lippmann Schwinger states. Given the growing interest in this alternative description, it is important that the  theoretical foundations of the effective equilibrium theory be more concretely established and details of the formulation refined. In this paper we provide a pedagogical exposition of the core concepts of the effective equilibrium theory. Furthermore, we introduce a novel perturbative framework for interacting systems, which makes explicit use of its Lippmann-Schwinger scatting states.

First, we discuss the effective equilibrium theory by making use of the open system limit \cite{Doyon_Andrei} which guarantees that the system relaxes to a steady state. This well-controlled mathematical construct mimics the role of relaxation mechanisms, and does not require the explicit inclusion of the bath degrees of freedom at the level of the Hamiltonian. We also demonstrate the equivalence between observables computed using the Schwinger-Keldysh framework and the effective equilibrium approach. It is important to emphasize the fact that the effective equilibrium description encompasses the case when one includes interactions on the dot.

Next, we outline a systematic scheme for computing Green's functions perturbatively in the interaction 
strength, which also provides the basis for an extension to 
non-perturbative analytical methods. Having established the basic methodology, we use the Anderson impurity model \cite{Anderson_1961} out of equilibrium as a prototype to 
illustrate the details of our perturbative approach. Working at the particle-hole symmetric point, we derive the self-energy of an electron on the dot to second order in the interaction strength, capturing the complete (non-linear) bias dependence. This result is then used to study the spectral function within the Born Approximation. The observed central peak can be regarded as a precursor of the much sharper Abrikosov-Suhl (Kondo) peak. Our results correctly reduce to the well-known Fermi liquid expansion in the limit of small bias, temperature and frequency \cite{Hewson,Nozieres_1974}. When increasing the bias voltage, there is a complete suppression of the Kondo effect 
and no resonance splitting is observed. We then break particle-hole symmetry by introducing a local magnetic field on the dot and analyze the effect of the bias and the Zeeman energy on the Abrikosov-Suhl resonance \cite{Muehlbacher}. This approximation 
is known to qualitatively reflect all features one expects in 
the low energy sector (in particular at low bias voltages 
we obtain a typical trident-shaped curve). 

Our paper is organized as follows. In Section 2, we introduce the model and define the ``open-system limit'' \cite{Mehta_Andrei_new,Doyon_Andrei}. This concept provides a transparent framework that guarantees the existence and uniqueness of a steady-state density matrix. In Section 3, we summarize the effective equilibrium approach introduced by Hershfield \cite{hershfield_reformulation_1993} and demonstrate the equivalence between observables computed within this framework and the Schwinger-Keldysh formalism. This enables us to arrive at the Meir-Wingreen formula for the current \cite{Meir_Wingreen_1,Meir_Wingreen_2}. We then derive a compact expression for the charge occupation on the dot, whereby we identify the spectral function of the dot as the central quantity of interest. In Section 4, we introduce a novel perturbative framework for treating interactions which captures the effect of the bias voltage non-perturbatively. In Section 5, we present our results for the current and charge occupancy on the dot for the Anderson impurity model \cite{Anderson_1961}. The details of the calculations are included in the appendices.

\section{Model}

\begin{figure}[h]
    \centering
        \includegraphics{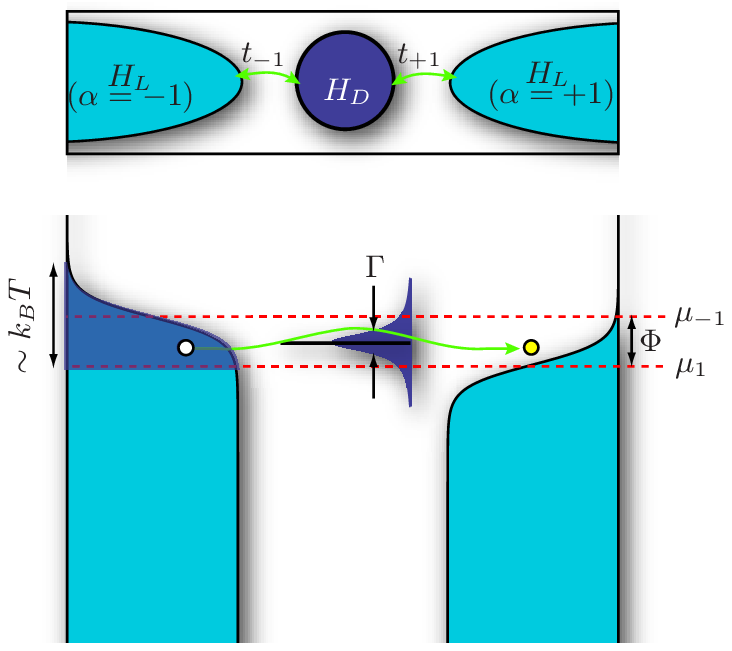}
        \caption{Schematic setup (upper panel) and energy diagram (lower panel) of a generic quantum impurity model out of equilibrium. The system is at a temperature $T$, and $\Phi=\mu_{1}-\mu_{-1}\equiv eV$ denotes the voltage bias between the source and drain leads. The (bare) energy of the dot level is given by $\epsilon_{d}$. The energy broadening of this level, given by the width $\Gamma$, is due to the tunnel coupling between dot and leads.\label{fig:transport}}    
    \vskip 0.2cm
\end{figure}

In our discussion of the effective equilibrium approach, we consider, for the sake of concreteness, a system consisting of two Fermi-liquid leads, coupled by tunnel junctions to a central system with a single energy level, the ``dot"  (see Fig.\ \ref{fig:transport}). The Hamiltonian of such a generic system can be written in the form $H=H_{L}+H_{D}+H_{T}$. The first term
\be
H_{L}=\sum_{\alpha k \sigma}\epsilon_{\alpha k}c^{\dag}_{\alpha k\sigma}c_{\alpha k\sigma}
\ee
describes the left and right leads ($\alpha=\pm1$), where $c_{\alpha k\sigma}$ ($c_{\alpha k\sigma}^\dag)$ annihilates (creates) an electron (strictly speaking a Fermi-liquid quasiparticle) in state $k$ with spin projection $\sigma$ in lead $\alpha$. The corresponding energy dispersion is denoted by $\epsilon_{\alpha k}$. These leads couple to the dot, whose Hamiltonian is given by
\be
H_{D}=\ \sum_{\sigma}\epsilon_{d}d^{\dag}_{\sigma}d_{\sigma}+H_\text{int},
\ee
where $d_{\sigma}$ ($d_{\sigma}^\dag$) annihilates (creates) an electron with spin $\sigma$ in the  discrete level with energy $\epsilon_d$. Any electron interactions are lumped into the contribution $H_\text{int}$, and we will assume that these interactions are localized on the dot (as it is the case for the Anderson model). Finally, the tunneling of electrons between leads and dot is captured by the tunneling Hamiltonian
\be
H_{T}=\frac{1}{\sqrt{\Omega}}\sum_{\alpha k \sigma}t_{\alpha k}\left(c^{\dag}_{\alpha k\sigma}d_{\sigma}+\text{h.c.}\right),
\label{eq:tunneling}
\ee
where $\Omega$ is the lead volume (assumed identical for both leads) and $t_{\alpha k}$ specifies the tunneling matrix element for electron transfer between state $k$ in lead $\alpha$ and the discrete dot state.
In the presence of a bias voltage, realized as a chemical potential difference  $\Phi=\mu_{1}-\mu_{-1}$ between left and right lead, the tunneling induces an electric current. 


Typically, the steady state is reached after a short time determined by relevant relaxation rates in the leads. For the purpose of calculations aiming at steady-state quantities, it is convenient to avoid the consideration of microscopic relaxation mechanisms and instead invoke the so-called ``open-system limit"  \cite{Mehta_Andrei_new,Doyon_Andrei}. In short, this approach proceeds as follows:
Initially, up to some time $t=t_0<0$ in the early past, the tunneling term is absent and the Hamiltonian of the system is given by $H_{0}=H_{L}+H_{D}$, such that leads and dot are decoupled. For $t<t_0$ the system is hence described by the separable density matrix
\begin{align}
	\rho_{0}=\exp\left[-\beta\left(H_{0}-\frac{\Phi}{2}\sum_{\alpha}\alpha N_{\alpha}\right)\right].
\end{align}
Here, $\beta=(k_B T)^{-1}$ denotes the inverse temperature and  $N_{\alpha}=\sum_{k\sigma}c^{\dag}_{\alpha k\sigma}c_{\alpha k\sigma}$ is the number operator of electrons in lead $\alpha$. Between times $t_0<t<0$, the tunneling is then `switched on' adiabatically, {\it i.e.}, \ $H=H_{0}+H_{T}e^{\eta t}\theta(t-t_{0})$, where the parameter $\eta\rightarrow 0^{+}$ defines a slow switch-on rate.  At time $t=0$ the tunneling has reached its full strength, the system is in its steady state and observables can be evaluated.

As demonstrated in Refs.\ \cite{Mehta_Andrei_new} and \cite{Doyon_Andrei}, the existence and uniqueness of the steady state is tied to the validity of the inequalities $v_F/L \ll |t_0|^{-1} \ll \eta$, where $v_F$ denotes the Fermi velocity and $L$ the linear system size. Intuitively, these inequalities ensure that hot electrons hopping onto a given lead at time $t_0$ will not be reflected back and return to the junction before the measurement process, and further that the process of switching on $H_T$ remains adiabatic. We also note that the energy scale of switch-on $|t_0|^{-1}$ suffices to smear out the energy level spacing $v_F/L$. In this sense, the openness of the system provides the ``dissipation'' mechanism necessary for the steady state, allowing the high-energy electrons to escape to infinity and thus, effectively relax.
On a more technical level, the inequalities result in the factorization of long-time correlation functions, which facilitates the proof of existence and uniqueness of steady state (see Appendix A and Appendix B). 
 
The crucial ingredient in our discussion is that $L/v_F$ determines the largest time scale in our problem. The exact protocol by which we switch on the tunneling is irrelevant for the formation of steady state. To illustrate this fact, take for example the non-interacting resonant level, and for convenience assume that the leads are identical ({\it i.e.}, $t_{\alpha k}=t$). Now, in stark contrast to the open-system limit we switch the tunneling on instantaneously. One then observes that the transients decay with a relaxation time $\sim1/\Gamma$, where 
$\Gamma=2\pi t^{2}\nu$ denotes the linewidth of the dot \cite{Thomas_2008}. Here $\nu$ symbolizes the density of states, which is assumed to be constant. To make the argument concrete, assume that the tunneling Hamiltonian is absent for $t<0$, and the distribution functions in the leads are given by Fermi functions $f(\ep-\al\frac{\Phi}{2})$. Furthermore, suppose that the dot is initially unoccupied, {\it i.e.}, $n_{d}^{\text{in}}=0$. The occupation of the dot at a time t($>0$) is given by \cite{Thomas_2008}
 \begin{align}
n_{d}(t)&=n_{d_{\text{ss}}}\left(1+e^{-2 \Gamma t}\right)\notag\\
&-\frac{2\Gamma e^{-\Gamma t}}{\pi}\int d\omega\left[f(\omega-\frac{\Phi}{2})+f(\omega+\frac{\Phi}{2})\right]\frac{\text{cos}\left[(\omega-\ep_{d})t\right]}{\Gamma^{2}+(\omega-\ep_{d})^{2}}.
\end{align}
Here
\begin{align}
n_{d_{\text{ss}}}=\frac{\Gamma}{\pi}\int d\omega\frac{f(\omega-\frac{\Phi}{2})+f(\omega+\frac{\Phi}{2})}{\Gamma^{2}+(\omega-\ep_{d})^{2}}
\label{eq:ssoccupation}
\end{align}
denotes the steady state expectation value of the dot occupation, after the transients have decayed. It is interesting to note that if we turn the coupling of the QD to the left and right leads to zero at the same rate, (formally this implies taking $\Gamma\rightarrow 0$ in Eq. (\ref{eq:ssoccupation})) then the final QD occupation is given by
\begin{align}
n_{d}^{\text{fin}}=f\left(\ep_{d}-\frac{\Phi}{2}\right)+f\left(\ep_{d}+\frac{\Phi}{2}\right).
\end{align} 
The fact that $n_{d}^{\text{in}}\neq n_{d}^{\text{fin}}$, clearly illustrates the irreversibility of the turning on process. Thus, the tunneling is treated exactly within the effective equilibrium approach. Interactions on the other hand can be turned on adiabatically and this procedure is completely reversible, and does not exhibit similar anomalies. It is thus possible to treat interactions perturbatively.

\section{Effective Equilibrium Theory}

The description of a nonequilibrium problem in terms of an effective equilibrium density matrix was first proposed by Hershfield in Ref.\ \cite{hershfield_reformulation_1993}. This Section is organized as follows. First, we outline the theoretical fundamentals of this equilibrium description. Then, we show the equivalence between observables computed within the Schwinger-Keldysh scheme and the effective equilibrium theory. In Appendices A and B, we flesh out the proof of the existence and uniqueness of the effective equilibrium density matrix, making careful use of the ``open-system limit'' \cite{Mehta_Andrei_new,Doyon_Andrei}.

\subsection{Effective Equilibrium Density Matrix}

The nonequilibrium steady-state density matrix $\rho$ is rewritten in the usual Boltzmann form,
\begin{equation}\label{rhoHershfield}
\rho=\exp\left[-\beta(H - Y)\right],
\end{equation}
at the cost of introducing an additional operator, which we will call (following Hershfield's convention) the $Y$ operator.\footnote{Note that throughout the text, density matrices are not assumed to be normalized. Normalization is always introduced explicitly when evaluating expectation values.} Formally, the definition of this correction operator as 
\be
Y=\frac{1}{\beta}\ln \rho+ H
\ee
 is always possible. However, in this form it is neither particularly elucidating nor useful for calculating nonequilibrium transport properties. Hershfield put forward the idea that $Y$ can be expressed explicitly and compactly in terms of Lippmann-Schwinger operators \cite{gell-mann_formal_1953,Lippmann}  
 $\psi_{\alpha k\sigma}$,
which are fermionic operators that diagonalize the full Hamiltonian \cite{Han_2006}
\begin{align}\label{eq:LS_ham}
H=\sum_{\alpha k}\epsilon_{\alpha k}\psi^{\dag}_{\alpha k\sigma}\psi_{\alpha k\sigma},
\end{align}
and can be formally expressed by the equation
\begin{equation}
\psi^\dagger_{\alpha k\sigma}=c^\dagger_{\alpha k\sigma}+\frac{1}{\epsilon_{\alpha k}-\mathcal{L}+i\eta}\mathcal{L}_{T} c^\dagger_{\alpha k\sigma}
\label{eq:lippmannformal}.
\end{equation}
The Liouvillian superoperators $\mathcal{L}$, $\mathcal{L}_{T}$ and $\mathcal{L}_{Y}$ are defined such that that their action on any operator $\mathcal{O}$ is given by $\mathcal{L}\mathcal{O}=[H,\mathcal{O}]$, $\mathcal{L}_{T}\mathcal{O}=[H_{T},\mathcal{O}]$ and $\mathcal{L}_{Y}\mathcal{O}=[Y,\mathcal{O}]$ respectively. 

Hershfield showed that the $Y$ operator has the general form \cite{hershfield_reformulation_1993}
\begin{equation}
Y=\frac{\Phi}{2}\sum_{\alpha k\sigma}\alpha\psi^{\dag}_{\alpha k\sigma}\psi_{\alpha k\sigma}.\label{Yrep}
\end{equation}
Since this operator encodes the entire $\Phi$ dependence, the $Y$ operator is also called the bias operator. Note that $Y$ vanishes at zero bias and the steady-state density matrix correctly simplifies to the equilibrium density matrix. 

In Appendices A and B, we provide a detailed derivation of $Y$ given in Eq.\ \eqref{Yrep}. In contrast to Hershfield, we do not invoke the presence of additional relaxation mechanisms to reach steady state. Instead, we make systematic use of the time-dependent open system approach  \cite{Mehta_Andrei_new,Doyon_Andrei}.

\subsection{Correspondence with the Schwinger-Keldysh Approach\label{sec:correspondence}}

We now illustrate how to compute transport observables such as the current and the charge occupation on the dot within this description. We propose an imaginary-time formulation for treating such nonequilibrium systems in a manner similar to finite temperature field theory.

\noindent 
In imaginary-time, we define the propagation of an operator by
\begin{align}
\mathcal{ O}(\tau)&=e^{\tau(H-Y)}\mathcal{ O}e^{-\tau(H-Y)}=e^{\tau(\mathcal{ L}-\mathcal{ L}_{Y})}\mathcal{ O}.
\end{align}
The nonequilibrium thermal Green's function is defined on $0<\tau<\beta$ as
\begin{align}
\mathcal{ G}_{\mathcal{ O}_{1}\mathcal{ O}_{2}}(\tau)=-\langle\mathcal{ T}\left[\mathcal{ O}_{1}(\tau)\mathcal{ O}_{2}(0)\right]\rangle=-\langle\mathcal{ O}_{1}(\tau)\mathcal{ O}_{2}(0)\rangle.
\end{align}
The subsequent Fourier transform in imaginary-time results in
\begin{align}
\mathcal{ G}_{\mathcal{ O}_{1}\mathcal{ O}_{2}}(i\omega_{n})=\left\langle\left\{\mathcal{ O}_{1},\frac{e^{i\omega_{n}0^{{+}}}}{i\omega_{n}-\mathcal{ L}+\mathcal{ L}_{Y}}\mathcal{ O}_{2}\right\}\right\rangle,
\end{align}
where $\omega_{n}=(2n+1)\pi/\beta$ ($n\in {\mathbb Z}$) denotes the fermionic Matsubara frequencies.

Switching to real-time, the Heisenberg representation of an operator $\mathcal{ O}$ is given by $\mathcal{ O}(t)=e^{iHt}\mathcal{ O}e^{-iHt}$. The nonequilibrium real-time retarded Green's function can then be expressed as 
\begin{align}
G^{\text{ret}}_{\mathcal{ O}_{1}\mathcal{ O}_{2}}(t)&=-i\theta(t)\langle\{\mathcal{ O}_{1}(t),\mathcal{ O}_{2}(0)\}\rangle\notag\\
&=-i\theta(t)\frac{\Tr\left[e^{-\beta(H-Y)}\left\{\mathcal{ O}_{1}(t),\mathcal{ O}_{2}(0)\right\}\right]}{\Tr\left[e^{-\beta(H-Y)}\right]}.
\end{align}
By using the spectral representation and then Fourier transforming, we obtain
\be
G^{\text{ret}}_{\mathcal{ O}_{1}\mathcal{ O}_{2}}(\omega)=\left\langle\left\{\mathcal{ O}_{1},\frac{1}{\omega-\mathcal{ L}+i\eta}\mathcal{ O}_{2}\right\}\right\rangle.
\ee
We emphasize that despite the effective equilibrium character of the Hershfield approach, one important difference between the effective equilibrium description and the traditional equilibrium many-body formalism remains. This distinction arises from the different propagators  in imaginary versus real-time; namely, in Hershfield's effective equilibrium formalism all Heisenberg operators in imaginary-time evolve under the modified Hamiltonian $H-Y$, whereas Heisenberg operators in real-time  evolve under the Hamiltonian $H$. As a result, imaginary-time and real-time Green's function, $\mathcal{ G}_{\mathcal{ O}_{1}\mathcal{ O}_{2}}(i\omega_{n})$ and $G^{\text{ret}}_{\mathcal{ O}_{1}\mathcal{ O}_{2}}(\omega)$  are not related via the analytic continuation $i\omega_{n}\rightarrow\omega+i\eta$.

As an observable of prime interest in transport, let us now consider the current flowing through the dot. It is given by
\begin{align}
I&=\frac{I_{1}+I_{-1}}{2}=-\frac{e}{2}\left\langle\frac{d\left(N_{1}(t)-N_{-1}(t)\right)}{dt}\right\rangle\notag\\
&=i\frac{e}{2}\sum_{\alpha}\alpha\left\langle[N_{\alpha}(t),H]\right\rangle=i\sum_{\alpha k\sigma}\alpha \frac{et_{\alpha k}}{2\sqrt{\Omega}}\left\langle\left(c^{\dag}_{\alpha k\sigma}d_{\sigma}-d^{\dag}_{\sigma}c_{\alpha k\sigma}\right)\right\rangle\notag\\
&=\iim\left[\sum_{\alpha k \sigma}\alpha\frac{et_{\alpha k}}{\sqrt{\Omega}}\mathcal{ G}_{c_{\alpha k \sigma}d_{\sigma}^\dag}(\tau=0)\right],
\label{current_primary}
\end{align}
where $-e$ denotes the charge of the electron.
Here, we have used the fact that the system is in a steady state and the current is unchanged under time translations. For a diagrammatic analysis, it is convenient to express the current in terms of the Fourier representation of the imaginary-time Green's function, namely
\begin{align}
 I&=\iim\left[\sum_{\alpha k \sigma\omega_{n}}\alpha \frac{e t_{\alpha k}}{\sqrt{\Omega}}\frac{1}{\beta}\mathcal{ G}_{c_{\alpha k \sigma}d_{\sigma}^\dag}(i\omega_{n})\right].
\label{eq:current1}
\end{align}

We now show the equivalence of this approach with the Schwinger-Keldysh formalism and recover the familiar Meir-Wingreen formula for the steady-state current \cite{Meir_Wingreen_1,Meir_Wingreen_2}. 
For simplicity, let us assume that the coupling to the dot is independent of $k$, \emph{i.e.} $t_{\alpha k}=t_{\alpha}$, and that the leads are identical $\epsilon_{\alpha k}=\epsilon_{k}$ (this is not a strict requirement and the proof can be easily generalized). Our starting point is Eq.\ \eqref{eq:current1} and we now write explicitly
\begin{align}
&\frac{1}{\beta}\sum_{\omega_{n}}\mathcal{ G}_{c_{\alpha k \sigma}d_{\sigma}^\dag}(i\omega_{n})=\frac{1}{\beta}\sum_{\omega_{n}}\left\langle\left\{c_{\alpha k \sigma},\frac{e^{i\omega_{n}0^{{+}}}}{i\omega_{n}-\mathcal{ L}+\mathcal{ L}_{Y}}d^{\dag}_{\sigma}\right\}\right\rangle.
\end{align}
Using Eq.\ \eqref{eq:lippmannformal} for $c_{\alpha k\sigma}$ and evaluating the effect of $\mathcal{L}_T$, we obtain
\begin{align}
&\frac{1}{\beta}\sum_{\omega_{n}}\mathcal{ G}_{c_{\alpha k \sigma}d_{\sigma}^\dag}(i\omega_{n})=\frac{1}{\beta}\sum_{\omega_{n}}\bigg[\left\langle\left\{\psi_{\alpha k \sigma},\frac{e^{i\omega_{n}0^{{+}}}}{i\omega_{n}-\mathcal{ L}+\mathcal{ L}_{Y}}d^{\dag}_{\sigma}\right\}\right\rangle\notag\\
&-\left\langle\left\{\frac{1}{\epsilon_{k}+\mathcal{ L}-i\eta}d_{\sigma},\frac{e^{i\omega_{n}0^{{+}}}}{i\omega_{n}-\mathcal{ L}+\mathcal{ L}_{Y}}d^{\dag}_{\sigma}\right\}\right\rangle\bigg].\label{aux1}
\end{align}

Making an additional observation, one can show that the second term does not need to be evaluated when computing the current $I$. To see this, we exploit the fact that in steady state, the current in the left and right junctions must be identical. Hence, we can write the current as a weighted average of the form  $I=\left(t_{1}^{2}I_{-1}+t_{-1}^{2}I_{1}\right)/\left(t_{1}^{2}+t_{-1}^{2}\right)$. This eliminates the presence of the second term in Eq.\ \eqref{aux1} from the expression for the current. Proceeding with the remaining term we obtain in several steps,
\begin{align}
\frac{1}{\beta}\sum_{\omega_{n}}\left\langle\left\{\psi_{\alpha k \sigma},\frac{e^{i\omega_{n}0^{{+}}}}{i\omega_{n}-\mathcal{ L}+\mathcal{ L}_{Y}}d^{\dag}_{\sigma}\right\}\right\rangle
&\ =\frac{t_{\alpha}}{\sqrt{\Omega}}f\left(\epsilon_{k}-\alpha\frac{\Phi}{2}\right)\left\langle\left\{\frac{1}{\epsilon_{k}+\mathcal{ L}-i\eta}d_{\sigma},d^{\dag}_{\sigma}\right\}\right\rangle\notag\\
&=\frac{t_{\alpha}}{\sqrt{\Omega}}f\left(\epsilon_{k}-\alpha\frac{\Phi}{2}\right) G^\text{ret}_{d_{\sigma}d_{\sigma}^\dag}(\epsilon_{k}),
\label{part1}
\end{align}
where the transition from the first to the second line is facilitated by the general relations summarized in Appendix C.

Using this identity together with the replacement $\frac{1}{\Omega}\sum_{k}\rightarrow \int \nu\,d\epsilon_{k}$ in Eq. \eqref{current_primary}, where $\nu$ denotes the density of states (assumed constant), we obtain the well-known Meir-Wingreen expression for the current \cite{Meir_Wingreen_1,Meir_Wingreen_2}
\begin{align}
I=&2e\frac{\Gamma_{1}\Gamma_{-1}}{\Gamma_{1}+\Gamma_{-1}}\int d\epsilon_{k}A_{d}(\epsilon_{k})\left[f\left({\epsilon_{k}+\frac{\Phi}{2}}\right)-f\left({\epsilon_{k}-\frac{\Phi}{2}}\right)\right].
\label{eq:meirwingreen}
\end{align}
Here,
\begin{align}
A_{d}(\epsilon_{k})&=-\frac{1}{\pi}\sum_{\sigma}\iim \left[G^{\text{ret}}_{d_{\sigma}d_{\sigma}^\dag}(\epsilon_{k})\right]
\label{eq:spectral_func}
\end{align}
 is the nonequilibrium spectral function of the dot and is in general a function of the bias voltage $\Phi$. However, in our notation we shall suppress this explicit bias dependence of the spectral function for convenience. In deriving Eq.\ \eqref{eq:spectral_func} we have adopted the standard procedure and linearized the spectrum about the Fermi surface. Furthermore, $\Gamma_{\alpha}=\pi t_{\alpha}^{2}\nu$ denotes the partial broadening of the level due to the coupling to the lead $\alpha$. 

Next, we derive a more general form of the Meir Wingreen formula. This is accomplished by the following sequence of steps.

Proceeding as in Eq.\ \eqref{aux1} one observes
\begin{align}
\frac{1}{\beta}\sum_{\omega_{n}}\mathcal{ G}_{d_{\sigma} c_{\alpha k \sigma}^\dag}(i\omega_{n})&=\frac{1}{\beta}\sum_{\omega_{n}}\bigg[\left\langle\left\{d_{\sigma},\frac{e^{i\omega_{n}0^{{+}}}}{i\omega_{n}-\mathcal{ L}+\mathcal{ L}_{Y}}\psi^{\dag}_{\alpha k \sigma}\right\}\right\rangle\notag\\
&\qquad\qquad-\left\langle\left\{\frac{e^{i\omega_{n}0^{{+}}}}{i\omega_{n}+\mathcal{ L}-\mathcal{ L}_{Y}}d_{\sigma},\frac{1}{\epsilon_{k}-\mathcal{ L}+i\eta}d^{\dag}_{\sigma}\right\}\right\rangle\bigg].\label{aux2}
\end{align}
Using a straightforward manipulation on the lines of Eq.\ \eqref{part1}, the first part of the expression in Eq.\ \eqref{aux2} can be simplified to give
\begin{align}
&\frac{1}{\beta}\sum_{\omega_{n}}\left\langle\left\{d_{\sigma},\frac{e^{i\omega_{n}0^{{+}}}}{i\omega_{n}-\mathcal{ L}+\mathcal{ L}_{Y}}\psi^{\dag}_{\alpha k \sigma}\right\}\right\rangle=\frac{t_{\alpha}}{\sqrt{\Omega}}f\left(\epsilon_{k}-\alpha\frac{\Phi}{2}\right)G^{\text{adv}}_{d_{\sigma}d^{\dag}_{\sigma}}(\epsilon_{k}).
\label{part2}
\end{align}
Thus, using Eq.\ \eqref{current_primary} together with Eqs.\ \eqref{part1} and \eqref{part2} we get 

\begin{align}
I&=i\sum_{\alpha k\sigma}\alpha \frac{et_{\alpha}}{2\sqrt{\Omega}}\left(\mathcal{ G}_{d_{\sigma}c^{\dag}_{\alpha k\sigma}}(\tau=0)-\mathcal{ G}_{c_{\alpha k\sigma}d^{\dag}_{\sigma}}(\tau=0)\right)\notag\\
&=i\frac{e}{2\pi}\sum_{\alpha\sigma}\alpha\Gamma_{\alpha}\int_{-\infty}^{\infty} d\epsilon_{k}\bigg\{f\left(\epsilon_{k}-\alpha\frac{\Phi}{2}\right)\left[G^{\text{adv}}_{d_{\sigma}d^{\dag}_{\sigma}}(\epsilon_{k})-G^{\text{ret}}_{d_{\sigma}d^{\dag}_{\sigma}}(\epsilon_{k})\right]\notag\\
&\quad\qquad-\frac{1}{\beta}\sum_{\omega_{n}}\left\langle\left\{\frac{e^{i\omega_{n}0^{{+}}}}{i\omega_{n}+\mathcal{ L}-\mathcal{ L}_{Y}}d_{\sigma},\frac{1}{\epsilon_{k}-\mathcal{ L}+i\eta}d^{\dag}_{\sigma}\right\}\right\rangle \notag\\
&\quad\qquad-\frac{1}{\beta}\sum_{\omega_n}
\left\langle\left\{\frac{1}{\epsilon_{k}+\mathcal{ L}-i\eta}d_{\sigma},\frac{e^{i\omega_{n}0^{{+}}}}{i\omega_{n}-\mathcal{ L}+\mathcal{ L}_{Y}}d^{\dag}_{\sigma}\right\}\right\rangle \bigg\}.
\end{align}

Using an identity from Appendix \ref{app:2}, and integrating over $\epsilon_{k}$ in the last 2 terms, we find
\begin{align}
I&=i\frac{e}{2\pi}\sum_{\alpha\sigma}\alpha\Gamma_{\alpha}\bigg\{\int_{-\infty}^{\infty} d\epsilon_{k} f\left(\epsilon_{k}-\alpha\frac{\Phi}{2}\right)\left[G^{\text{adv}}_{d_{\sigma}d^{\dag}_{\sigma}}(\epsilon_{k})-G^{\text{ret}}_{d_{\sigma}d^{\dag}_{\sigma}}(\epsilon_{k})\right] \notag\\
&\qquad\qquad\quad\qquad+2i\pi \mathcal{ G}_{d_{\sigma}d^{\dag}_{\sigma}}(\tau=0)\bigg\}.
\end{align}
Now, $2i\pi \mathcal{ G}_{d_{\sigma}d^{\dag}_{\sigma}}(\tau=0)=2i\pi\left\langle d^{\dag}_{\sigma}d_{\sigma}\right\rangle=2\pi G^{<}_{d_{\sigma}d^{\dag}_{\sigma}}(t=0)=\int_{-\infty}^{\infty}d\epsilon_{k} G^{<}_{d_{\sigma}d^{\dag}_{\sigma}}(\epsilon_{k})$. Thus, we finally recover the more general form of the Meir-Wingreen formula:
\begin{align}
I&=i\frac{e}{2\pi}\sum_{\alpha\sigma}\alpha\Gamma_{\alpha}\int_{-\infty}^{\infty} d\epsilon_{k} \bigg\{f\left(\epsilon_{k}-\alpha\frac{\Phi}{2}\right)\left[G^{\text{adv}}_{d_{\sigma}d^{\dag}_{\sigma}}(\epsilon_{k})-G^{\text{ret}}_{d_{\sigma}d^{\dag}_{\sigma}}(\epsilon_{k})\right]+ G^{<}_{d_{\sigma}d^{\dag}_{\sigma}}(\epsilon_{k})\bigg\}.
\end{align}

We proceed in a similar fashion to derive an expression for the charge occupation on the dot (See Appendix C for details)
\begin{align}
n_{d}&=\sum_{\sigma}\langle d^{\dag}_{\sigma}d_{\sigma}\rangle=\frac{1}{\beta}\sum_{\sigma,\omega_{n}}\mathcal{ G}_{d_{\sigma}d_{\sigma}^\dag}(i\omega_{n})\notag\\
&=\frac{1}{2}\int d\epsilon_{k} A_{d}(\epsilon_{k})\left[f\left(\epsilon_{k}-\frac{\Phi}{2}\right)+f\left(\epsilon_{k}+\frac{\Phi}{2}\right)\right]\notag\\
&=\int d\epsilon_{k} A_{d}(\epsilon_{k})f^{\text{eff}}(\epsilon_{k},\Phi),
\label{eq:occupation}
\end{align}
where 
\begin{align}
f^{\text{eff}}(\epsilon_{k},\Phi)=\frac{1}{2}\left[f\left(\epsilon_{k}-\frac{\Phi}{2}\right)+f\left(\epsilon_{k}+\frac{\Phi}{2}\right)\right].
\end{align}

Note that on taking the limit $\Phi\rightarrow 0$, Eq. \eqref{eq:occupation} reduces to its well-known form from equilibrium many-body theory. In Appendix\ \ref{app:2} we provide a straightforward derivation of this result using the spectral representation within the effective equilibrium formulation. This result can also be obtained within the Keldysh framework as we have shown in Appendix\ \ref{app:3}.   

We thus identify the non-equilibrium spectral function 
$ A_{d}(\omega)$ as the central quantity of interest, which one can use to calculate transport observables. In the effective equilibrium scheme this can be done directly, without resorting to the Keldysh contour which leads to Dyson equations which couple the different Green's functions. 

\subsection{Simple Application: Non-Interacting Resonant Level}

\label{NIRLM}
The situation in which interactions on the dot are absent ({\it i.e.}, $H_{\text{int}}=0$) corresponds to the double-barrier problem in quantum mechanics and presents the simplest scenario, and has been solved exactly via numerous approaches \cite{Landauer,Buttiker}. We shall recall the key results for this model outlining the methodology we adopt for obtaining them within the effective equilibrium framework. In Section 4 we shall use these results as the starting point for perturbative calculations for interacting theories.

The formal expression for the Lippmann-Schwinger \cite{gell-mann_formal_1953,Lippmann} operators is given by

\begin{equation}
\psi^\dagger_{\alpha k\sigma}=c^\dagger_{\alpha k\sigma}+\frac{1}{\epsilon_{\alpha k}-\mathcal{L}+i\eta}\mathcal{L}_{T} c^\dagger_{\alpha k\sigma}.
\end{equation}
For the case of the non-interacting resonant level we shall use the notation $\psi^{(0)}_{\alpha k\sigma}$ to denote the Lippmann-Schwinger operators in order to distinguish them from the Lippmann-Schwinger operators of interacting models.
It is straightforward to show that
\begin{align}
\psi^{(0)\dag}_{\alpha k\sigma}&=c^{\dag}_{\alpha k\sigma}+\frac{t}{\sqrt{\Omega}}g_{d}(\epsilon_{k})d^{\dag}_{\sigma}+\frac{t^{2}}{\Omega}\sum_{\alpha'k'}\frac{g_{d}(\epsilon_{k})}{\epsilon_{k}-\epsilon_{k'}+i\eta}c^{\dag}_{\alpha'k'\sigma}
\label{eq:LS_NIRLM}
\end{align}
follows immediately from Eq.\ \eqref{eq:lippmannformal}. Here, we have defined
\begin{align}
g_{d}(\epsilon_{k})=\frac{1}{\epsilon_{k}-\epsilon_{d}+i\Gamma},
\label{eq:gd}
\end{align}
and $\Gamma=\Gamma_{1}+\Gamma_{-1}$ denotes the electronic level broadening of the  
dot.

Eq. \eqref{eq:LS_NIRLM} can be readily inverted to express $\{ c\}$ and $\{ d\}$ in terms of $\{\psi^{(0)}\}$ to give
\be
d_{\sigma}^{\dag}=\frac{t}{\sqrt{\Omega}}\sum_{\alpha k}g^{\ast}_{d}(\epsilon_{k})\psi_{\alpha k\sigma}^{(0)\dag}.
\label{eq:d expansion}
\ee

\be
c^\dagger_{\alpha k\sigma}=\psi^{(0)\dagger}_{\alpha k\sigma}-
\frac{t^{2}}{\Omega}\sum_{\alpha'k'}\frac{g^{\ast}_{d}(\epsilon_{k'})}{\epsilon_{k}-\epsilon_{k'}+i\eta}\psi_{\alpha'k'\sigma}^{(0)\dag}
\label{eq:c expansion}.
\ee

In the imaginary-time formalism we use Eq.\ \eqref{eq:meirwingreen} to compute the current
\begin{align}
I&= -\frac{e t}{\sqrt{\Omega}}\frac{1}{\beta}\iim\left[\sum_{\alpha k\sigma\omega_{n}}\alpha\mathcal{ G}_{d_{\sigma}c_{\alpha k\sigma}^\dag}(i\omega_{n})\right].
\end{align}
Using Eqs. \ \eqref{eq:d expansion} and \eqref{eq:c expansion}, the current can be finally written as
\begin{align}
I=&-\frac{e t^{2}}{\Omega}\frac{1}{\beta}\iim\left[\sum_{\alpha k\sigma\omega_{n}}\sum_{\alpha'k'}\alpha g_{d}(\epsilon_{k'})\mathcal{ G}_{\psi^{(0)}_{\alpha'k'\sigma} \psi^{(0)\dag}_{\alpha k\sigma}}(i\omega_{n})\right].
\label{eq:current2}
\end{align}

Note that the expression for the current, in terms of the imaginary-time Green's function $\mathcal{ G}_{\psi^{(0)}_{\alpha'k'\sigma} \psi^{(0)\dag}_{\alpha k\sigma}}(i\omega_{n})$ is exact even when including interactions on the dot. In this context however, the density matrix is no longer diagonal in terms of $\{\psi^{(0)}\}$, the Lippmann-Schwinger states of the non-interacting resonant level. Instead, they are diagonalized by the  Lippmann-Schwinger operators of the interacting model $\{\psi\}$. 

The charge occupation on the dot is given by
\begin{align}
n_{d}&=\sum_{\sigma}\langle d^{\dag}_{\sigma}d_{\sigma}\rangle=\frac{1}{\beta}\sum_{\sigma,\omega_{n}}\mathcal{ G}_{d_{\sigma}d_{\sigma}^\dag}(i\omega_{n}).
\label{eq:occupation2}
\end{align}

For the non-interacting resonant level the Green's functions in terms of Lippmann-Schwinger operators are trivial to evaluate, since the action in the generating functional is quadratic and diagonal in terms of these states and hence
\begin{align}
\mathcal{ G}_{\psi^{(0)}_{\alpha'k'\sigma'} \psi^{(0)\dag}_{\alpha k\sigma}}(i\omega_{n})=-\frac{e^{i\omega_{n}0^{{+}}}}{-i\omega_{n}+\epsilon_{k}-\alpha\frac{\Phi}{2}}\delta_{\sigma\sigma'}\delta_{\alpha\alpha'}\delta_{kk'}.
\end{align}

Using this expression in Eq.\ \eqref{eq:current2} gives
\begin{align}
I=\ &\frac{e\Gamma^{2}}{\pi}\int_{-\infty}^{\infty}\frac{1}{(\epsilon-\epsilon_{d})^2+\Gamma^2}\left[f\left(\epsilon+\frac{\Phi}{2}\right)-f\left(\epsilon-\frac{\Phi}{2}\right)\right]d\epsilon.
\label{eq:NIRLMcurrent}
\end{align}

Comparison of Eq.\ (\ref{eq:NIRLMcurrent}) with Eq.\ \eqref{eq:meirwingreen} allows one to extract the nonequilibrium spectral function
\begin{align}
A_{d}(\epsilon)=\frac{2\Gamma}{\pi}\frac{1}{(\epsilon-\epsilon_{d})^2+\Gamma^2}.
\label{eq:spectrum_NIRLM}
\end{align}
It is instructive to note that for the non-interacting case the spectral function is independent of the bias. The introduction of interactions generally imparts to the spectral function a nontrivial bias dependence, as exemplified by the Anderson model which we shall discuss in Section 5.

The electron occupation on the dot is evaluated in a similar fashion, namely
\begin{align}
\label{eq:NIRLMoccupation}
n_{d}=\ &\frac{1}{2}\int_{-\infty}^{\infty}A_{d}(\epsilon)\left[f\left(\epsilon-\frac{\Phi}{2}\right)+f\left(\epsilon+\frac{\Phi}{2}\right)\right]d\epsilon\notag\\
=\ &\frac{\Gamma}{\pi}\int_{-\infty}^{\infty}\frac{1}{(\epsilon-\epsilon_{d})^{2}+\Gamma^{2}}\left[f\left(\epsilon-\frac{\Phi}{2}\right)+f\left(\epsilon+\frac{\Phi}{2}\right)\right]d\epsilon.
\end{align}

In the noninteracting case it is possible to express the bias operator $Y$ explicitly in terms of the electron operators in the leads and dot,
\begin{align}
Y&=\sum_{\alpha k\sigma}\frac{\alpha\Phi}{2}\psi^{\dag}_{\alpha k\sigma}\psi_{\alpha k\sigma}\notag\\
&=\sum_{\alpha k\sigma}\frac{\alpha\Phi}{2}\bigg[c^{\dag}_{\alpha k\sigma}c_{\alpha k\sigma}+\frac{t}{\sqrt{\Omega}}\left(g_{d}(\epsilon_{k})d_{\sigma}^{\dag}c_{\alpha k\sigma}+\text{h.c.}\right)\notag\\
&\qquad\qquad\qquad+\frac{t^{2}}{\Omega}\sum_{\alpha'k'}\left(\frac{g_{d}(\epsilon_{k})}{\epsilon_{k}-\epsilon_{k'}+i\eta}c^{\dag}_{\alpha'k'\sigma}c_{\alpha k\sigma}+\text{h.c.}\right)\bigg], 
\label{eq:Yoperator_original}
\end{align}
which is far more intricate than the chemical potential term in the initial density matrix $Y_{0}=\sum_{\alpha k\sigma}\frac{\alpha\Phi}{2}c^{\dag}_{\alpha k\sigma}c_{\alpha k\sigma}$, since it encodes the entire nonequilibrium boundary condition of the system, and the bias voltage manifests itself as off-diagonal contributions for the electrons in the leads and the dot. This can be attributed to the highly non-local nature of the scattering states. 

\section{Perturbation Theory for Interacting Models}
\label{PT}
In this Section we develop a novel perturbative framework for interacting theories which facilitates the use of well-established many body techniques.

The Section is structured as follows. In Subsection 4.1 we obtain a formal expansion of the Lippmann-Schwinger operators of the interacting system $\{\psi\}$ in terms of the Lippmann-Schwinger operators of the non-interacting resonant level $\{\psi^{(0)}\}$, in powers of the interaction strength.  This expansion is used in Subsection 4.2 to  explicitly construct the {\it effective Hamiltonian}, and hence the density operator. In Subsection 4.3 we lay the framework for computing real and imaginary-time Green's functions, perturbatively in the strength of the interaction, using the effective equilibrium density operator of Subsection 4.2.

A key feature of this formalism is that it is an effective equilibrium theory. It bypasses the need to define Green's functions on the Keldysh contour and eliminates the necessity to invoke coupled Dyson's equations. However, in contrast to the Keldysh formalism, the {\it effective Hamiltonian}  is not available a priori and needs to be explicitly constructed to the desired order.

It is important to note that the scheme presented in this Section is a model independent generic procedure, barring the fact that the interactions are assumed to be confined to the vicinity of dot.  In Section 5, we use the Anderson model as a prototype to illustrate the details of this scheme.

\subsection{Construction of Lippmann-Schwinger operators}
In this Subsection we sketch the formal expansion of the  Lippmann- Schwinger operators, in powers of the interaction strength, using a convenient starting point. Typically the point of reference is an exactly solvable model, for example the non-interacting level, the Toulouse point of the Kondo model or the infinite-U Anderson model in the $N\rightarrow\infty$ limit  ($N$ being the number of ``flavors'' of the electrons). For the purpose of the ensuing discussion we use the non-interacting resonant level as our point of reference and construct the  Lippmann- Schwinger operators in terms of $\{\psi^{(0)}\}$ in powers of the interaction strength.  

We introduce the additional superoperators $\mathcal{L}'$ and $\mathcal{L_{I}}$ such that their action on any operator $\mathcal{O}$ is given by $\mathcal{L}'\mathcal{O}=[H-H_{\text{int}},\mathcal{O}]$ and $\mathcal{L}_{I}\mathcal{O}=[H_{\text{int}},\mathcal{O}]$ respectively. Using Eq. \eqref{eq:lippmannformal} we obtain
\begin{align}
\psi^\dagger_{\alpha k\sigma}&=c^\dagger_{\alpha k\sigma}
+\frac{1}{\epsilon_{k}-\mathcal{ L}+i\eta}\mathcal{ L}_T 
c^\dagger_{\alpha k\sigma}\notag\\
&=c^\dagger_{\alpha k\sigma}
+\left[1-\frac{1}{\epsilon_{k}-\mathcal{ L'}+i\eta}\mathcal{ L}_{I}\right]^{-1}\frac{1}{\epsilon_{k}-\mathcal{ L'}+i\eta}\mathcal{ L}_T 
c^\dagger_{\alpha k\sigma}\notag\\
&=c^\dagger_{\alpha k\sigma}
+\sum_{l=0}^{\infty}\left[\frac{1}{\epsilon_{k}-\mathcal{ L'}+i\eta}\mathcal{ L}_{I}\right]^{l}\frac{1}{\epsilon_{k}-\mathcal{ L'}+i\eta}\mathcal{ L}_T 
c^\dagger_{\alpha k\sigma}.
\label{eq:temp1}
\end{align}

Since the Lippmann-Schwinger operators for the non-interacting resonant level satisfy
\begin{equation}
{\psi}^{(0)\dagger}_{\alpha k\sigma}=c^\dagger_{\alpha k\sigma}
+\frac{1}{\epsilon_{k}-\mathcal{ L'}+i\eta}\mathcal{ L}_T 
c^\dagger_{\alpha k\sigma},
\end{equation}
we can rewrite Eq.\ \eqref{eq:temp1} as 
\begin{align}
\psi^\dagger_{\alpha k\sigma}&=\psi^{(0)\dagger}_{\alpha k\sigma}
+\sum_{l=1}^{\infty}\left[\frac{1}{\epsilon_{k}-\mathcal{ L'}+i\eta}\mathcal{ L}_{I}\right]^{l}(\psi^{(0)\dagger}_{\alpha k\sigma}-c^\dagger_{\alpha k\sigma}).
\end{align}
This expression can be simplified to give
\begin{align}
\psi^\dagger_{\alpha k\sigma}=\ &\psi^{(0)\dagger}_{\alpha k\sigma}
+\frac{t^{2}}{\Omega}\sum_{l=1}^{\infty}\left[\frac{1}{\epsilon_{k}-\mathcal{ L'}+i\eta}\mathcal{ L}_{I}\right]^{l}
\sum_{\alpha'k'}\frac{g^{\ast}_{d}(\epsilon_{\alpha'k'})}{\epsilon_{k}-\epsilon_{k'}+i\eta}\psi^{(0)\dagger}_{\alpha'k'\sigma}.
\end{align}
This can be expressed symbolically as 
\begin{align}
\psi^\dagger_{\alpha k\sigma}\equiv \sum_{l=0}^{\infty}\psi^{\dag(l)}_{\alpha k\sigma},
\label{eq:LS_operator_expansion}
\end{align}
where the superscript $l$ denotes the contribution proportional to the $l$-th power of $H_{\text{int}}$.

The formal expansion in Eq.\ \eqref{eq:LS_operator_expansion} is used in the following subsection to construct the density matrix order by order in the interaction strength.

\subsection{Construction of the Density Matrix}
The steady state density matrix given by Eq. \eqref{rhoHershfield} is of the form
\begin{align}
\rho=e^{-\beta(H-Y)}\equiv e^{-\beta\mathcal{ H}},
\end{align}
where $\mathcal{ H}=H-Y$ denotes the {\it effective Hamiltonian} for the system. The effective Hamiltonian is diagonal in terms of the Lippmann-Schwinger operators and can be written as 
\begin{align}
&\mathcal{ H}=\sum_{\al k\sigma}\left(\ep_{k}-\al\frac{\Phi}{2}\right)\psi^\dagger_{\alpha k\sigma}\psi_{\alpha k\sigma}.
\label{effH}
\end{align}
Our objective is to construct the effective Hamiltonian, and thereby the density matrix, in powers of the interaction strength. This will be instrumental for computing observables via nonequilibrium Green's functions, as we shall illustrate in Subsection 4.3.
Inserting Eq.\ \eqref{eq:LS_operator_expansion} in Eq. \eqref{effH} above, we get
\begin{align}
\mathcal{ H}&=\sum_{\al k\sigma}\left(\ep_{k}-\al\frac{\Phi}{2}\right)\sum_{n_{1},n_{2}=0}^{\infty}\psi^{\dagger(n_{1})}_{\alpha k\sigma}\psi^{(n_{2})}_{\alpha k\sigma}\notag\\
&=\sum_{m=0}^{\infty}\left[\sum_{\al k\sigma}\left(\ep_{k}-\al\frac{\Phi}{2}\right)\sum_{p=0}^{m}\psi^{\dagger(p)}_{\alpha k\sigma}\psi^{(m-p)}_{\alpha k\sigma}\right]\equiv\sum_{m=0}^{\infty}\mathcal{ H}^{(m)}.
\label{eq:eff_Ham}
\end{align}
Here,
\begin{align}
\mathcal{ H}^{(m)}=\left[\sum_{\al k\sigma}\left(\ep_{k}-\al\frac{\Phi}{2}\right)\sum_{p=0}^{m}\psi^{\dagger(p)}_{\alpha k\sigma}\psi^{(m-p)}_{\alpha k\sigma}\right]
\label{eq:rho_expansion}
\end{align}
denotes the contribution to the effective Hamiltonian $\mathcal{ H}$ proportional to $\left(H_{\text{int}}\right)^{m}$. Note that the $m=0$ term,
\begin{align}
\mathcal{ H}^{(0)}=\sum_{\al k\sigma}\left(\ep_{k}-\al\frac{\Phi}{2}\right)\psi^{(0)\dagger}_{\alpha k\sigma}\psi^{(0)}_{\alpha k\sigma},
\end{align}
is precisely the effective Hamiltonian for the non-interacting resonant level.

We emphasize the fact that the perturbative scheme developed here is very distinct from the usual perturbative expansion familiar from equilibrium quantum field theory. In the present approach, the ``perturbation'' is not available {\it a priori}, rather the effective Hamiltonian is to be computed to the desired order in the interaction strength before resorting to diagrammatic methods.

\subsection{Perturbative computation of Green's functions}
Next, we use the effective Hamiltonian from Eq.\ \eqref{eq:eff_Ham} to compute Green's functions to any desired order in the interaction strength. We present two distinct frameworks for computations, in real and imaginary-time respectively. 

The imaginary-time formulation involves a straightforward implementation of the effective Hamiltonian via a functional integral representation. This is achieved by a coherent state representation (over Grassman variables) of the Lippmann-Schwinger operators of the non-interacting resonant level. The motivation behind the use of this representation is the fact that these operators are simply related to the $\{c\}$ and $\{d\}$ operators, as given by Eqs.\ \eqref{eq:d expansion} and \eqref{eq:c expansion}. All transport quantities are formulated in terms of the  $\{c\}$ and $\{d\}$ operators which represent the original degrees of freedom of the system. 

For the $\{\psi^{(0)}\}$ operators, the generic single-particle Green's function is given by
\begin{equation*}
\mathcal{ G}_{\psi^{(0)}_{\alpha_{1}k_{1}\sigma_{1}}\psi^{(0)\dag}_{\alpha_{2}k_{2}\sigma_{2}}}(\tau)=-\left\langle\mathcal{ T}\left[\psi^{(0)}_{\alpha_{1}k_{1}\sigma_{1}}(\tau)\psi^{(0)\dag}_{\alpha_{2}k_{2}\sigma_{2}}(0)\right]\right\rangle.
\end{equation*}
Note, the density matrix used in computing the above expectation value is diagonal in terms of the Lippmann-Schwinger states of the interacting model. We apply the formal expansion for $\mathcal{ H}$ from Eq.\ \eqref{eq:rho_expansion} in the above expression, in order to rewrite the effective Hamiltonian in terms of the Lippmann-Schwinger states of the non-interacting resonant level retaining terms to the desired order in $H_{\text{int}}$.  This is implemented by expressing the partition function as a coherent state  functional integral

\begin{align}
\mathcal{ G}_{\psi^{(0)}_{\alpha_{1}k_{1}\sigma_{1}}\psi^{(0)\dag}_{\alpha_{2}k_{2}\sigma_{2}}}(\tau)&=\ \frac{1}{Z}\int \prod_{\alpha k\sigma} (\mathcal D \psi^{(0)\ast}_{\alpha k \sigma} \mathcal D \psi^{(0)}_{\alpha k\sigma})\psi^{(0)\ast}_{\alpha_{2}k_{2}\sigma_{2}}(0)\psi^{(0)}_{\alpha_{1}k_{1}\sigma_{1}}(\tau)\times\notag\\
&\exp\left[-\int_{0}^{\beta}d\tau'\sum_{\alpha k\sigma}\bigg\{\psi^{(0)\ast}_{\alpha k\sigma}(\tau')\partial_{\tau'}\psi^{(0)}_{\alpha k\sigma}(\tau')+\mathcal{ H}\left(\psi^{(0)\ast},\psi^{(0)}\right)\bigg\}\right].
\end{align}
Here, the partition function is given by
\begin{align}
Z=&\Tr\left[e^{-\beta\mathcal{ H}}\right]=\int \prod_{\alpha k\sigma} (\mathcal D \psi^{(0)\ast}_{\alpha k \sigma} \mathcal D \psi^{(0)}_{\alpha k\sigma})\notag\\
&\times\exp\left[-\int_{0}^{\beta}d\tau'\sum_{\alpha k\sigma}\bigg\{\psi^{(0)\ast}_{\alpha k\sigma}(\tau')\partial_{\tau'}\psi^{(0)}_{\alpha k\sigma}(\tau')+\mathcal{ H}\left(\psi^{(0)\ast},\psi^{(0)}\right)\bigg\}\right],
\label{Z}
\end{align}
and we have suppressed explicit time labels for the Grassmann variables featuring in the effective Hamiltonian. Using Eq.\ \eqref{eq:rho_expansion} in the exponent of the above equation we can formally write

\begin{align}
\mathcal{ G}_{\psi^{(0)}_{\alpha_{1}k_{1}\sigma_{1}}\psi^{(0)\dag}_{\alpha_{2}k_{2}\sigma_{2}}}(\tau)&=\ \frac{1}{Z}\int \prod_{\alpha k\sigma} (\mathcal D \psi^{(0)\ast}_{\alpha k \sigma} \mathcal D \psi^{(0)}_{\alpha k\sigma})\psi^{(0)\ast}_{\alpha_{2}k_{2}\sigma_{2}}(0)\psi^{(0)}_{\alpha_{1}k_{1}\sigma_{1}}(\tau)
\notag\\
\times\exp\bigg[-\int_{0}^{\beta}d\tau'&\sum_{\alpha k\sigma}\bigg\{\psi^{(0)\ast}_{\alpha k\sigma}(\tau')\partial_{\tau'}\psi^{(0)}_{\alpha k\sigma}(\tau')+\sum_{m=0}^{\infty}\mathcal{H}^{(m)}\left(\psi^{(0)\ast},\psi^{(0)}\right)\bigg\}\bigg].
\end{align}
 Introducing the free action $S'= \int_{0}^{\beta}d\tau'\sum_{\alpha k\sigma}\left\{\psi^{(0)\ast}_{\alpha k\sigma}(\tau')\partial_{\tau'}\psi^{(0)}_{\alpha k\sigma}(\tau')+\mathcal{ H}^{(0)}\left(\psi^{(0)\ast},\psi^{(0)}\right)\right\}$ and using the abbreviated notation $ \mathcal{ H}^{(m)}\left(\psi^{(0)\ast},\psi^{(0)}\right)=\mathcal{ H}^{(m)}(\tau')$, we can expand the exponent to obtain

\begin{align}
&\mathcal{ G}_{\psi^{(0)}_{\alpha_{1}k_{1}\sigma_{1}}\psi^{(0)\dag}_{\alpha_{2}k_{2}\sigma_{2}}}(\tau)=\ \frac{1}{Z}\int \prod_{\alpha k\sigma} (\mathcal D \psi^{(0)\ast}_{\alpha k \sigma} \mathcal D \psi^{(0)}_{\alpha k\sigma})\psi^{(0)\ast}_{\alpha_{2}k_{2}\sigma_{2}}(0)\psi^{(0)}_{\alpha_{1}k_{1}\sigma_{1}}(\tau)e^{-S'}
\notag\\
&\qquad\qquad\qquad\qquad\times\bigg\{\sum_{\nu=0}^{\infty}\frac{(-1)^{\nu}}{\nu!}\prod_{i=1}^{\nu}\int_{0}^{\beta}d\tau_{i}\bigg(\sum_{m=1}^{\infty}\mathcal{ H}^{(m)}(\tau_{i})\bigg)\bigg\}\notag\\
&\qquad=\ \frac{1}{Z}\int \prod_{\alpha k\sigma} (\mathcal D \psi^{(0)\ast}_{\alpha k \sigma} \mathcal D \psi^{(0)}_{\alpha k\sigma})\psi^{(0)\ast}_{\alpha_{2}k_{2}\sigma_{2}}(0)\psi^{(0)}_{\alpha_{1}k_{1}\sigma_{1}}(\tau)e^{-S'}\times
\notag\\
&\qquad\sum_{n=0}^{\infty}\bigg\{\sum_{l=1}^{n}\frac{(-1)^{l}}{l!}\sum_{m_{1}+\ldots+m_{l}=n}\left(\int_{0}^{\beta}d\tau_{1}\mathcal{H}^{(m_{1})}(\tau_{1})\right)\ldots\left(\int_{0}^{\beta}d\tau_{l}\mathcal{H}^{(m_{l})}(\tau_{l})\right)\bigg\}.
\label{eq:convention}
\end{align}
In the last step, we have reorganized powers of $H_{\text{int}}$ by introducing the indices $ m_{1},m_{2},\ldots,m_{l}\in\{1,2,...\}$ and used the convention that the term in curly brackets is unity for $n=0$. This allows us to isolate the order $n$ contribution. Thus we can write
\begin{align}
\mathcal{ G}_{\psi^{(0)}_{\alpha_{1}k_{1}\sigma_{1}}\psi^{(0)\dag}_{\alpha_{2}k_{2}\sigma_{2}}}(\tau)=\sum_{n=0}^{\infty}\mathcal{ G}^{(n)}_{\psi^{(0)}_{\alpha_{1}k_{1}\sigma_{1}}\psi^{(0)\dag}_{\alpha_{2}k_{2}\sigma_{2}}}(\tau),
\end{align}
where the $n$-th order contribution is explicitly given by
\begin{align}
\mathcal{G}^{(n)}_{\psi^{(0)}_{\alpha_{1}k_{1}\sigma_{1}}\psi^{(0)\dag}_{\alpha_{2}k_{2}\sigma_{2}}}&(\tau)=\ \frac{1}{Z}\int\prod_{\alpha k\sigma} (\mathcal D \psi^{(0)\ast}_{\alpha k \sigma} \mathcal D \psi^{(0)}_{\alpha k\sigma})\psi^{(0)\ast}_{\alpha_{2}k_{2}\sigma_{2}}(0)\psi^{(0)}_{\alpha_{1}k_{1}\sigma_{1}}(\tau)e^{-S'}\times
\notag\\
\sum_{l=1}^{n}\bigg\{\frac{(-1)^{l}}{l!}&\sum_{m_{1}+\ldots+m_{l}=n}\left(\int_{0}^{\beta}d\tau_{1}\mathcal{H}^{(m_{1})}(\tau_{1})\right)\ldots\left(\int_{0}^{\beta}d\tau_{l}\mathcal{H}^{(m_{l})}(\tau_{l})\right)\bigg\}.
\label{eq:imag_GF}
\end{align}

The formal expression in Eq. \eqref{eq:imag_GF} above, captures the contribution proportional to $\left(H_{\text{int}}\right)^{n}$ to the single-particle Green's function. This procedure can be used to compute Green's functions, and hence observables, perturbatively. The imaginary-time formulation enables one to use the  gamut of functional methods and may also be used for nonperturbative analyses, {\it e.g.} the infinite-$U$ Anderson model in the limit of $N\rightarrow\infty$.

We now illustrate the procedure for computing Green's functions in real-time.  In this case the expansion is more intricate, since it involves a double expansion in powers of $H_{\text{int}}$. In Section 3.2 we established the retarded electron Green's function of the dot as the quantity of central importance in the real-time description. This can be expressed in the frequency domain as 
\begin{align}
G^{\text{ret}}_{d_{\sigma}d^{\dag}_{\sigma}}(\omega)=\left\langle \left\{\frac{1}{\omega+\mathcal{ L}+i\eta}d_{\sigma},d^{\dag}_{\sigma}\right\}\right\rangle=\frac{\Tr \left[e^{-\beta\mathcal{ H}}\left\{\frac{1}{\omega+\mathcal{ L}+i\eta}d_{\sigma},d^{\dag}_{\sigma}\right\}\right]}{\Tr\left[e^{-\beta\mathcal{ H}}\right]}.
\end{align}
Note that the bias operator appears in both the exponent and the Liouvillian superoperator. To compute the above expression perturbatively, we systematically collect powers of $U$ arising from the expansion of both the exponent and also the Liouvillian. First, we expand $\frac{1}{\omega+\mathcal{ L}+i\eta}d_{\sigma}$ in powers of $U$
\begin{align}
D_{\sigma}(\omega)&\equiv\frac{1}{\omega+\mathcal{ L}+i\eta}d_{\sigma}=\sum_{n_{1}=0}^{\infty}\left(\frac{-1}{\omega+\mathcal{ L}'+i\eta}\mathcal{ L_{\text I}}\right)^{n_{1}}\frac{1}{\omega+\mathcal{ L}'+i\eta}d_{\sigma}\notag\\
&\equiv\sum_{n_{1}=0}^{\infty}D^{(n_{1})}_{\sigma}(\omega),
\label{eq:Dn}
\end{align}
which implies the recursion relation
\begin{align}
D^{(n_{1})}_{\sigma}(\omega)=-\frac{1}{\omega+\mathcal{ L}'+i\eta}\mathcal{ L_{\text I}}D^{(n_{1}-1)}_{\sigma}(\omega).
\label{eq:recursion}
\end{align}

For any observable $\mathcal{O}$ the expectation value can be expanded in powers of the interaction as
\begin{align}
\langle\mathcal{ O}\rangle&=\frac{\Tr\left[e^{-\beta \mathcal{ H}}\mathcal{ O}\right]}{\Tr\left[e^{-\beta \mathcal{ H}}\right]}\notag\\
=&\sum_{\nu=0}^{\infty}\Tr\bigg[e^{-\beta \mathcal{ H}^{(0)}}\mathcal{ T}\bigg\{\frac{(-1)^\nu}{\nu!}\prod_{i=1}^{\nu}\int_{0}^{\beta}d\tau_{i}\bigg(\sum_{m=1}^{\infty}\mathcal{ H}^{(m)}(\tau_{i})\bigg)\bigg\}\mathcal{ O}\bigg]_{\text{conn}}\notag\\
=&\sum_{n_{2}=0}^{\infty}\Tr\bigg[ e^{-\beta \mathcal{ H}^{(0)}}\sum_{l=1}^{n_{2}}\mathcal{ T}\bigg\{\frac{(-1)^l}{l!}\prod_{i=1}^{l}\int_{0}^{\beta}d\tau_{i}\notag\\
&\times\bigg(\sum_{m_{1}+\ldots+m_{l}=n_{2}}\mathcal{ H}^{(m_{1})}(\tau_{1})\ldots\mathcal{ H}^{(m_{\nu})}(\tau_{l})\bigg)\bigg\}\mathcal{ O}\bigg]_{\text{conn}}\equiv\sum_{n_{2}=0}^{\infty}\langle\mathcal{ O}\rangle^{(n_{2})},
\end{align}
where $m_{1},m_{2},\ldots,m_{l}\in\{1,2,...\}$. Here, $\langle\mathcal{ O}\rangle^{n_{2}}$ represents the $n_{2}$-th order contribution to the expectation value of the observable. We use the convention given below Eq.\ \eqref{eq:convention} for the sums.  The subscript ``conn'' implies we discard all ``disconnected'' diagrams. Once the perturbation expansion is systematically implemented we use the standard diagrammatic rules for computing the contributions of the individual terms of the expansion. Finally, we systematically collect powers of $H_{\text{int}}$ arising from the two distinct  expansions
\begin{align}
G^{\text{ret}}_{d_{\sigma}d^{\dag}_{\sigma}}(\omega)&=\left\langle \left\{D_{\sigma}(\omega),d^{\dag}_{\sigma}\right\}\right\rangle=\sum_{n_{1},n_{2}=0}^{\infty}\left\langle \left\{D^{(n_{1})}_{\sigma}(\omega),d^{\dag}_{\sigma}\right\}\right\rangle^{(n_{2})}.
\end{align}
We rearrange the double sum and group together terms containing identical powers of $H_{\text{int}}$ to obtain
\begin{align}
G^{\text{ret}}_{d_{\sigma}d^{\dag}_{\sigma}}(\omega)&=\sum_{n=0}^{\infty}\left[\sum_{k=0}^{n}\left\langle \left\{D^{(k)}_{\sigma}(\omega),d^{\dag}_{\sigma}\right\}\right\rangle^{(n-k)}\right]\notag\\
&\equiv\sum_{n=0}^{\infty}G^{\text{ret}(n)}_{d_{\sigma}d^{\dag}_{\sigma}}(\omega).
\label{eq:retarded_ex}
\end{align}
Here, $G^{\text{ret}(n)}_{d_{\sigma}d^{\dag}_{\sigma}}$ denotes the $n$-th order contribution to the retarded electron Green's function of the dot.

We have thus constructed an explicit perturbative expansion for imaginary-time and real-time Green's functions applicable to generic interacting quantum impurity models. When compared to the Schwinger-Keldysh approach, this scheme allows a direct computation of the Green's function and bypasses the need for the Keldysh contour.

\section{Anderson Impurity Model} 
\label{AM}
In this Section, we use the Anderson impurity model as a prototype to implement the perturbative framework presented in Section 4. The Anderson impurity model \cite{Anderson_1961} describes a commonly encountered experimental scenario and serves as the microscopic model for a wide variety of fundamental physical phenomena. The interaction is given by
\begin{align}
H_{\text{int}}=\frac{U}{2}\widehat{n}_{d}\left(\widehat{n}_{d}-1\right)=\frac{U}{2}\sum_{\sigma}d^{\dag}_{\sigma}d^{\dag}_{-\sigma}d_{-\sigma}d_{\sigma},
\end{align} 
and denotes the Coulomb interaction between electrons of opposite spin projections on the quantum dot.  Here, $\widehat{n}_{d}=\sum_{\sigma}d^{\dag}_{\sigma}d_{\sigma}$ denotes the electron number operator of the dot. Thus, the interaction can be rewritten as 
\begin{align}
H_{\text{int}}=\frac{U}{2}\left(\frac{t^{2}}{\Omega}\right)^{2}\sum_{1,2,3,4,\sigma}g^{\ast}_{1}g^{\ast}_{2}g_{3}g_{4}\psi^{(0)\dag}_{1\sigma}\psi^{(0)\dag}_{2-\sigma}\psi^{(0)}_{3-\sigma}\psi^{(0)}_{4\sigma},
\label{eq:abb}
\end{align}
where we have introduced the abbreviated notation $l\equiv(\al_{l}k_{l})$ and $g_{l}\equiv g_{d}(\ep_{k_{l}})$, implying a sum over the lead index $\alpha$ and the quantum number $k$. Furthermore, we define $\sum_{1}\equiv\sum_{\alpha_{1}k_{1}}$ for the subsequent discussion.

Our analysis of the Anderson impurity model is organized as follows. We begin with a brief discussion of the mean field results of the Anderson model which follows immediately from the results obtained for the noninteracting resonant level in Section 3.3. In Subsection 5.2 we calculate the self energy to first order in the interaction strength within the imaginary-time framework and compare it with the mean field results. In Subsection 5.3 we study the nonequilibrium dynamics of the Anderson impurity at the particle-hole symmetric point within the second-order Born approximation. We also consider deviations from the particle-hole symmetric point by introducing a local magnetic field on the dot. 

\subsection{Mean Field Theory}
As a preliminary analysis we treat the Anderson model out of equilibrium within the mean-field approximation. This amounts to replacing the interaction Hamiltonian above with 
\begin{align}
H^\text{MF}_\text{int}=U\sum_{\sigma}\langle \hat{n}_{-\sigma}\rangle d^{\dag}_{\sigma}d_{\sigma}.
\end{align}
Here, $\hat{n}_{\sigma}=d^{\dagger}_{\sigma} d_{\sigma}$ gives the electron number operator of the dot for spin projection $\sigma$. At the mean-field level,  the interaction renormalizes the dot energy $\epsilon_{d}\rightarrow\left(\epsilon_{d}+U\langle \hat{n}_{-\sigma}\rangle\right)$, such that
\begin{align}
H^\text{MF}_{D}=\sum_{\sigma}\left(\epsilon_{d}+U\langle \hat{n}_{-\sigma}\rangle\right)d^{\dag}_{\sigma}d_{\sigma}.
\end{align}

Using Eq. \eqref{eq:NIRLMcurrent}, the expression for the current reads
\begin{align}
I=&\frac{e\Gamma^{2}}{2\pi}\sum_{\sigma}\int_{-\infty}^{\infty}\frac{1}{(\epsilon-\epsilon_{d}-U\langle \hat{n}_{-\sigma}\rangle)^2+\Gamma^2}\left[f\left(\epsilon+\frac{\Phi}{2}\right)-f\left(\epsilon-\frac{\Phi}{2}\right)\right]d\epsilon,
\label{eq:andersonMFT}
\end{align}
where the occupation number of the dot is determined self-consistently \cite{Zarand} via
\begin{align}
\langle \hat{n}_{\sigma}\rangle=\ &\frac{\Gamma}{2\pi}\int_{-\infty}^{\infty}\frac{1}{(\epsilon-\epsilon_{d}-U\langle \hat{n}_{-\sigma}\rangle)^2+\Gamma^2}\left[f\left(\epsilon+\frac{\Phi}{2}\right)+f\left(\epsilon-\frac{\Phi}{2}\right)\right]d\epsilon.
\end{align}
One observes that the condition for particle-hole symmetry, {\it i.e.}, $\langle \hat{n}_{\sigma}\rangle =1/2$ is realized for $\epsilon_d=-U/2$, as confirmed by first order in perturbation theory. 

\subsection{First-order analysis}
\begin{figure}[h]
\centering
\includegraphics[width=8.0cm]{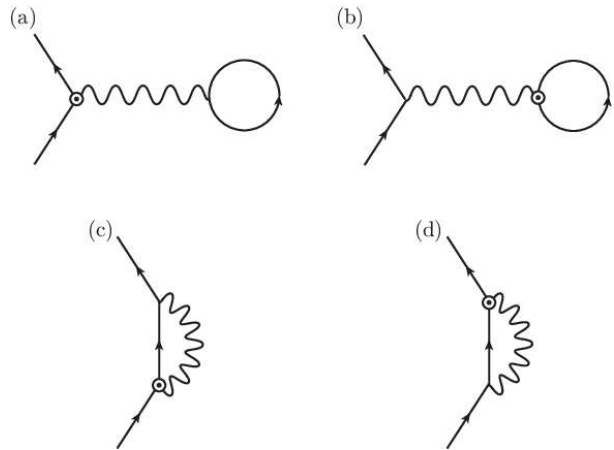}
\caption{Here, (a) and (b) denote Hartree diagrams contributing to the current to first-order in the interaction strength. The Fock diagrams (c) and (d) are irrelevant due to the exclusion principle. The absence of exchange symmetry is depicted by distinguishing a particular end of the interaction line by a dotted circle.}
\label{fig_1}
\end{figure}

The first order contribution to the Lippmann-Schwinger operator, cf. Eq. \eqref{eq:LS_operator_expansion}, is given by
\begin{align}
\psi^{\dag(1)}_{\alpha k\sigma}&=U g_{d}(\ep_{k})\left(\frac{t^{2}}{\Omega}\right)^{2}\sum_{123}\frac{g^{\ast}_{1}g^{\ast}_{2}g_{3}}{\ep_{k}-\ep_{1}-\ep_{2}+\ep_{3}+i\eta}\psi^{(0)\dag}_{1\sigma}\psi^{(0)\dag}_{2-\sigma}\psi^{(0)}_{3-\sigma}.
\end{align}
Using this expression in conjunction with Eq.\eqref{eq:rho_expansion} we construct the effective Hamiltonian to first-order in the interaction
\begin{align}
\mathcal{ H}^{(1)}&=\sum_{\alpha k\sigma}\left(\ep_{k}-\frac{\al\Phi}{2}\right)\left[\psi^{\dag(1)}_{\al k\sigma}\psi^{(0)}_{\al k\sigma}+\psi^{(0)\dag}_{\al k\sigma}\psi^{(1)}_{\al k\sigma}\right]\notag\\
&=\sum_{121'2'\sigma}\left(12\left|\mathcal{ V}\right|1'2'\right)\psi^{(0)\dag}_{1\sigma}\psi^{(0)\dag}_{2-\sigma}\psi^{(0)}_{2'-\sigma}\psi^{(0)}_{1'\sigma},
\end{align}

where we have defined the matrix element
\begin{align}
\left(12\left|\mathcal{ V}\right|1'2'\right)&=U\left(\frac{t^{2}}{\Omega}\right)^{2}\left[\ep_{1'}-\ep_{1}-\left(\al_{1'}-\al_{1}\right)\frac{\Phi}{2}\right]\frac{g_{1}^{\ast}g_{2}^{\ast}g_{1'}g_{2'}}{\ep_{1'}+\ep_{2'}-\ep_{1}-\ep_{2}+i\eta}.
\end{align}
Note that $\left(12\left|\mathcal{ V}\right|1'2'\right)$ is not symmetric under $(1\leftrightarrow2,1'\leftrightarrow2')$. In diagrammatic language, one needs to distinguish opposite ends of the (two-body) interaction line, as we have illustrated in Fig.\ \ref{fig_1}. 

Assuming identical leads, we use Eq. \eqref{eq:current2} to compute the first order correction to the current (using standard diagrammatic techniques),
\begin{align}
I^{(1)}=&-\frac{e t^{2}}{\Omega}\frac{1}{\beta}\iim\left[\sum_{11'\omega_{n}}\alpha g_{1'}\mathcal{G}^{(1)}_{\psi^{(0)}_{1'\sigma} \psi^{(0)\dagger}_{1\sigma}}(i\omega_{n})\right]\notag\\
=&\ Un_{d}^{(0)}\left[\frac{e\Gamma^{2}}{\pi}\int d\ep_{1}
\frac{\left(\ep_{1}-\ep_{d}\right)\left[f\left(\ep_{1}-\frac{\Phi}{2}\right)-f\left(\ep_{1}+\frac{\Phi}{2}\right)\right]}{\left[(\ep_{1}-\ep_{d})^{2}+\Gamma^{2}\right]\left[(\ep_{1}-\ep_{d})^{2}+\Gamma^{2}\right]}\right],
\label{eq:first_order_new}
\end{align}

where
\begin{align}
n_{d}^{(0)}=\frac{2\Gamma}{\pi}\int d\ep \frac{f^{\text{eff}}\left(\ep,\Phi\right)}{\left[(\ep-\ep_{d})^{2}+\Gamma^{2}\right]}
\end{align}
denotes the charge occupation of the dot for the noninteracting resonant level as given by Eq. \eqref{eq:NIRLMoccupation}.

Note, for the first order computation only the Hartree diagrams contribute whereas the Fock diagrams are irrelevant. Comparing the expression for the current in Eq.\ \eqref{eq:first_order_new} with Eq.\ \eqref{eq:meirwingreen} we extract the spectral function to first order in the interaction  
\begin{align}
A_{d}(\ep)=\frac{2\Gamma}{\pi}\left[\frac{1}{\left[(\ep-\ep_{d})^{2}+\Gamma^{2}\right]}+U n_{d}^{(0)}\frac{\left[\ep-\ep_{d}\right]}{\left[(\ep-\ep_{d})^{2}+\Gamma^{2}\right]^{2}}\right].
\end{align}

One observes that this result agrees precisely with the mean field solution discussed above, to first-order in $U$, in accordance with Hartree-Fock theory.

\subsection{Dynamics about particle-hole symmetric point}

Next, we study quantum transport through an Anderson impurity which is at the particle-hole symmetric point, given by $\ep_{d}=-\frac{U}{2}$. The special feature of the particle-hole symmetric point is that by tuning the interaction strength $U$, one can pass smoothly from the weak to the strong coupling regime such that the Fermi liquid fixed point of the Hamiltonian remains invariant. Also, the Anderson impurity model maps exactly onto the Kondo model in the limit $U\rightarrow\infty$ at this point in parameter space.

In the following discussion it is convenient to absorb the bare energy of the dot in the interaction term, {\it i.e.}, we let $\ep_{d}\rightarrow\tilde{\ep}_{d}=\ep_{d}+\frac{U}{2}$, thereby generating an additional term $-\frac{U}{2}\sum_{\sigma}d^{\dag}_{\sigma}d_{\sigma}$. The particle-hole symmetric point is thus specified by $\tilde{\ep}_{d}=0$. This redefinition assures that the pole of the bare propagator is correctly positioned, and becomes particularly important when we extend our results to the strong coupling regime. The interaction term is consequently redefined as
\begin{align}
H_{\text{int}}=\frac{U}{2}\left(\widehat{n}_{d}-1\right)^{2}.
\end{align}

Only even powers of $H_{\text{int}}$ feature in the perturbative expansion of the spectral function at the particle-hole symmetric point, and the leading correction is of order $\mathcal{O}\left(U^{2}\right)$. In our discussion, we adopt the real-time approach to compute the spectral function $A_{d}(\omega)$. In this context the real-time approach is computationally simpler than the imaginary-time approach since the $n$-th order calculation in $U$ requires us to compute the effective Hamiltonian only to order $n-1$, at the cost of evaluating $D _{\sigma}^{(n)}(\omega)$, cf. Eq. \eqref{eq:Dn}.

The non-interacting part of the Green's function is identified to be
\begin{align}
G^{\text{ret}(0)}_{d_{\sigma}d^{\dag}_{\sigma}}(\omega)=g_{d}(\omega),
\label{eq:order_zero}
\end{align}
which immediately followed from  Eq.\ \eqref{eq:retarded_ex}.
Evaluating the second order contribution we obtain (see Appendix \ref{app:second_order} for details)
\begin{align}
G^{\text{ret}(2)}_{d_{\sigma}d^{\dag}_{\sigma}}(\omega)&=U^{2}[g(\omega)^2]\left(\frac{\Gamma}{\pi}\right)^{3}\int d\ep_{1}\int d\ep_{2}\int d\ep_{3}\left|g_{1}\right|^{2}\left|g_{2}\right|^{2}\left|g_{3}\right|^{2}\notag\\
&\qquad\qquad\times\frac{\left[(1-f^{\text{eff}}_{1})f^{\text{eff}}_{2}f^{\text{eff}}_{3}+f^{\text{eff}}_{1}(1-f^{\text{eff}}_{2})(1-f^{\text{eff}}_{3})\right]}{\omega+\ep_{1}-\ep_{2}-\ep_{3}+i\eta},
\label{eq:2ndorder}
\end{align}
where we have abbreviated $f^{\text{eff}}_{n}=\frac{1}{2}\left[f\left(\ep_{n}-\frac{\Phi}{2}\right)+f\left(\ep_{n}+\frac{\Phi}{2}\right)\right]$.

The Dyson equation for the retarded electron Green's function of the dot is given by
\begin{align}
G^{\text{ret}}_{d_{\sigma}d^{\dag}_{\sigma}}(\omega)=G^{\text{ret}(0)}_{d_{\sigma}d^{\dag}_{\sigma}}(\omega)+G^{\text{ret}(0)}_{d_{\sigma}d^{\dag}_{\sigma}}(\omega)\Sigma_{\sigma}^{\text{ret}}(\omega)G^{\text{ret}}_{d_{\sigma}d^{\dag}_{\sigma}}(\omega).
\label{Dyson1}
\end{align}
Here, $\Sigma_{\sigma}^{\text{ret}}(\omega)$ is the proper self-energy obtained exclusively from contributions of one-particle irreducible diagrams. The Dyson equation above can be rewritten in the alternative form
\begin{align}
G^{\text{ret}}_{d_{\sigma}d^{\dag}_{\sigma}}(\omega)=G^{\text{ret}(0)}_{d_{\sigma}d^{\dag}_{\sigma}}(\omega)+G^{\text{ret}(0)}_{d_{\sigma}d^{\dag}_{\sigma}}(\omega)\bar{\Sigma}_{\sigma}^{\text{ret}}(\omega)G^{\text{ret}(0)}_{d_{\sigma}d^{\dag}_{\sigma}}(\omega),
\label{complete_sigma}
\end{align}
where $\bar{\Sigma}_{\sigma}^{\text{ret}}(\omega)$ represents the {\it complete} self-energy of the system, obtained by summing the 1-particle irreducible (1PI) diagrams.

Expanding $G^{\text{ret}}_{d_{\sigma}d^{\dag}_{\sigma}}(\omega)$ and $\bar{\Sigma}_{\sigma}^{\text{ret}}(\omega)$ in powers of $H_{\text{int}}$ it is possible to relate the $n$-th order ($n\geq1$) contribution to the self-energy 
\begin{align}
G^{\text{ret}(n)}_{d_{\sigma}d^{\dag}_{\sigma}}(\omega)=G^{\text{ret}(0)}_{d_{\sigma}d^{\dag}_{\sigma}}(\omega)\bar{\Sigma}_{\sigma}^{\text{ret}(n)}(\omega)G^{\text{ret}(0)}_{d_{\sigma}d^{\dag}_{\sigma}}(\omega).
\end{align}

Since the first order result is zero, we conclude that $\Sigma^{\text{ret}(2)}_{\sigma}=\bar{\Sigma}_{\sigma}^{\text{ret}(2)}$. The second-order contribution to the self-energy is directly read off from Eq.\ \eqref{eq:2ndorder}, and can be simplified to give
\begin{align}
\Sigma_{\sigma}^{\text{ret}(2)}(\omega)=U^{2}\left[\frac{1}{\omega+3i\Gamma}-3\left(\frac{\Gamma}{\pi}\right)^{2}\int{d\ep_{1}}\int{d\ep_{2}}\left|g_{1}\right|^{2}\left|g_{2}\right|^{2}g(\omega+\ep_{1}-\ep_{2})f_{1}^{\text{eff}}f_{2}^{\text{eff}}\right].
\label{eq:2nd_order_GF}
\end{align}

The low-bias behavior for the entire range of the parameter $U$ is captured in the Fermi liquid expansion \cite{Nozieres_1974} for the non-equilibrium spectral function and the differential conductance. This is typically expressed in terms of the susceptibilities $\tilde{\chi}_{\uparrow\downarrow}$ and $\tilde{\chi}_{\uparrow\uparrow}$ as \cite{Nozieres_1974} 
\begin{align}
A_{d}(\omega)&=\frac{2}{\pi\Gamma}\bigg[1-\left(\tilde{\chi}_{\uparrow\uparrow}^2+\frac{1}{2}\tilde{\chi}_{\uparrow\downarrow}^2\right)\left(\frac{\omega}{\Gamma}\right)^{2}-\frac{1}{2}\tilde{\chi}_{\uparrow\downarrow}^2\left(\frac{\pi T}{\Gamma}\right)^{2}-\frac{3}{8}\tilde{\chi}_{\uparrow\downarrow}^2\left(\frac{\Phi}{\Gamma}\right)^{2}+\ldots\bigg],
\label{eq:FL1}
\end{align}
and
\begin{align}
G&=G_{0}\bigg[1-\frac{\tilde{\chi}_{\uparrow\uparrow}^2+2\tilde{\chi}_{\uparrow\downarrow}^2}{3}\left(\frac{\pi T}{\Gamma}\right)^{2}-\frac{\tilde{\chi}_{\uparrow\uparrow}^2+5\tilde{\chi}_{\uparrow\downarrow}^2}{4}\left(\frac{\Phi}{\Gamma}\right)^{2}\ldots\bigg].
\label{eq:FL2}\end{align}
Here, $G=e\frac{dI}{d\Phi}$ denotes the differential conductance, and $G_{0}=2e^2/h$ is the conductance quantum. 

The susceptibilities $\tilde{\chi}_{\uparrow\downarrow}$ and $\tilde{\chi}_{\uparrow\uparrow}$ are analytic functions of $U/(\pi\Gamma)$ and their analytic properties have been extensively studied. It is well-known that their Taylor expansions display a rapid convergence \cite{Zvlatic}. Extracting the spectral function from the expression for the Green's function to second-order in Eq.\ \eqref{eq:2nd_order_GF}, one obtains
\begin{align}
A_{d}^{(2)}(\omega)&=\frac{2}{\pi\Gamma}\bigg[1-\left\{1 + \left(\frac{13}{2} - \frac{\pi^2}{2}\right) \left(\frac{U}{\pi\Gamma}\right)^{2}\right\}\left(\frac{\omega}{\Gamma}\right)^{2}-\notag\\
&\qquad\qquad\qquad\qquad\frac{1}{2}\left(\frac{U}{\pi\Gamma}\right)^{2}\left(\frac{\pi T}{\Gamma}\right)^{2}-\frac{3}{8}\left(\frac{U}{\pi\Gamma}\right)^{2}\left(\frac{\Phi}{\Gamma}\right)^{2}+\ldots\bigg],
\end{align}
and similarly
\begin{align}
G^{(2)}&=G_{0}\bigg[1-\left\{\frac{1}{3} + \frac{16 - \pi^2}{6} \left(\frac{U}{\pi\Gamma}\right)^2\right\}\left(\frac{\pi T}{\Gamma}\right)^{2}\notag\\
&\qquad\qquad\qquad\qquad-\left\{\frac{1}{4} + \frac{22 - \pi^2}{8} \left(\frac{U}{\pi\Gamma}\right)^{2}\right\}\left(\frac{\Phi}{\Gamma}\right)^{2}\ldots\bigg].
\end{align}

This agrees with Eqs.\ \eqref{eq:FL1} and \eqref{eq:FL2} where we use the well-known perturbative expansions for the susceptibilities $\tilde{\chi}_{\uparrow\downarrow}=-\frac{U}{\pi\Gamma}+\mathcal{ O}\left(\frac{U}{\pi\Gamma}\right)^{3}$ and $\tilde{\chi}_{\uparrow\uparrow}=1 + \frac{1}{2} \left(6 - \frac{\pi^2}{2}\right) \left(\frac{U}{\pi\Gamma}\right)^{2}+\mathcal{ O}\left(\frac{U}{\pi\Gamma}\right)^{3}$.

Strictly speaking, the expression for the  self-energy to order $U^{2}$ is a well-controlled approximation to the actual self-energy only when $U\ll\Gamma$. On the other hand Kondo physics becomes relevant in the regime $U\gg\Gamma$. For a qualitative analysis of this regime we apply the Born approximation using the second-order self energy obtained above \cite{Oguri_2002}.

%
%

It is in general difficult to rigorously justify the exact bounds of validity of the Born approximation without identifying an explicit control parameter, but this scheme is seen to give physical results for fairly large values of $U$($\sim 5\Gamma$) at the particle-hole symmetric point \cite{Oguri_2002}.

When $U\gg \Gamma$ and $\Phi\ll \Gamma$, we notice the appearance of the Abrikosov-Suhl resonance centered at $\omega=0$ reminiscent of the Kondo resonance, which corresponds to the case $U\rightarrow\infty$. This many-body resonance needs to be distinguished from the sidebands corresponding to the bare atomic levels which continue to persist. The resonance at the Fermi level diminishes with an increase in the bias voltage as shown in Fig. 3. 
One may conjecture that the effect of finite voltage might be captured by an effective temperature; this is in some sense related to a dephasing rate induced by
a finite current \cite{Kaminski}. Our results corroborate the absence of a two-channel Kondo effect
that would be associated with the coupling to two fermionic continua at different chemical potentials \cite{Wingreen,Rosch2}.
At large voltages the resonance is completely destroyed and one obtains the residual atomic sidebands at $\ep_{d}\pm\frac{U}{2}$. Comparing Fig. 3 and Fig. 4 we also confirm that the central peak has a diminishing width when increasing $U$. 

This second-order result, which is exact in the bias voltage, reduces in the low-bias and high-bias limits to those in similar discussion using the
Schwinger-Keldysh scheme \cite{Oguri_2002,Oguri_2001,Oguri_2007}. Qualitatively they agree with the NRG results and are confirmed by recent experiments \cite{Anders_2008,Leturcq}. The splitting of the Kondo resonance at intermediate biases is still an open issue since other groups have observed a split in the Abrikosov-Suhl resonance with increasing bias using a fourth-order calculation \cite{Fuji}. However, recent numerical calculations in the intermediate coupling regime \cite{Anders_2008,Millis,Muehlbacher,Han_2010,Thomas_2008,Thomas_2007} indicate that the peak splitting from the fourth-order perturbation theory has been overestimated.
  
Finally, we study deviations of the system from the particle-hole symmetric point. For this purpose we introduce a  local magnetic field ($H$) on the dot which implies $\tilde\epsilon_{d}\rightarrow \tilde\epsilon_{d}+\delta_{\sigma}$, thereby breaking particle-hole symmetry at $\tilde\epsilon_{d}=0$. Here, the deviation $\delta_{\sigma}=-\sigma\frac{H}{2}$ is given in the units defined by $g\mu_{b}=1$, where $g$ is the electron g-factor and $\mu_{b}$ denotes the Bohr magneton. The retarded electron Green's function of the dot for the non-interacting level now differentiates between the different spin projections
\begin{align}
G^{\text{ret}(0)}_{d_{\sigma}d^{\dag}_{\sigma}}(\omega)=\frac{1}{\omega-\delta_{\sigma}+i\Gamma}.
\end{align}

Away from the particle-hole symmetric point certain diagrammatic contributions to the self-energy which earlier evaluated to zero, cease to do so. The leading order contribution to the self-energy is at first-order. Using Eq.\ \eqref{eq:retarded_ex} we have computed the self-energy to second-order in the interaction, with details summarized in Appendix \ref{app:deviation}. The expression for the self-energy correctly captures the entire bias and field dependence to a given order in $U$. Using this result to approximate the retarded electron Green's function of the dot we observe a split in the Abrikosov-Suhl resonance as we increase the field strength $H$ as shown in Fig. 5. This split becomes discernible when $H\sim\Gamma$. Furthermore, the sidebands shift further apart as expected. For the Kondo model, which corresponds to the extreme limit $U\rightarrow\infty$, the effect of a magnetic field has been rigorously studied \cite{Prushke,Bauer_Oguri,Bauer_Koller}.	

The current in Fig. 6 is obtained via Eq. \eqref{eq:meirwingreen}, which at $T=0$ corresponds to integrating the spectral function over a window of length $\Phi$ centered on the dot. Note that the spectral function is a function of the bias and we observe suppression of the Abriksov-Suhl resonance with increasing bias accompanied by the atomic sidebands shifting further apart. In the $\Phi\rightarrow\infty$ limit the spectral function is integrated over the entire range of energies, which by  the sum rule $\int d\omega A_{d}(\omega)=1$ assures that the current saturates to the same asymptotic value. We see suppression of the current with increasing $U$ for low biases corresponding to a sharpening of the Abriksov-Suhl resonance. The sudden increase of the current for higher $\Phi$ corresponds to the contribution of the sidebands. 

%
 
\section{Conclusion}
In this paper, first we have discussed the theoretical foundations of the effective equilibrium theory initiated by Hershfield \cite{hershfield_reformulation_1993}, and demonstrated a correspondence with the observables computed using the Schwinger-Keldysh formalism. We have pedagogically re-derived the effective equilibrium density matrix approach proposed by Hershfield using the concept of an ``open system'', which does not require inclusion of additional relaxation mechanisms. Furthermore, we have illustrated the methodology of this formulation for a quantum dot, in which the reservoir leads play the role of (infinite) thermal baths, such that a unique steady state exists. The latter can be cast into the form of an effective equilibrium density matrix, where the associated modified Hamiltonian can be explicitly written in terms of the Lippmann-Schwinger scattering state operators. 

We also introduced a systematic perturbative framework for interacting systems out of equilibrium within the effective equilibrium theory. This procedure involves a novel expansion in terms of the Lippmann-Schwinger states of the system, in a manner very distinct from the Schwinger-Keldysh formalism. Using this method the need for the Keldysh contour is eliminated and the Green's function can be computed directly. This allows use of the usual Feynman diagrams albeit with a modified Hamiltonian. Furthermore, the perturbation theory is free of unphysical infrared divergences. However, the structure of the effective interactions becomes more complex with increasing order in perturbation theory.

We have illustrated our scheme with an application to the Anderson impurity subject to a local magnetic field and/or a finite bias. The non-equilibrium electron spectral function is evaluated analytically to second-order in the interaction strength, while capturing the entire bias and magnetic field dependence. We use the second order Born approximation for the self-energy to simulate the strong-coupling (Kondo) regime. The fate of the Abrikosov-Suhl resonance as a function of the bias and magnetic field was inferred from this result. We observed that increasing the bias causes a gradual decrease in the resonance until ultimately only the two atomic sidebands remain, indicating that the Kondo peak is destroyed with bias voltage without a perceptible splitting of the resonance peak. This result reduces in the low-bias and high-bias limits to that in similar discussion using the Schwinger-Keldysh scheme \cite{Oguri_2002,Oguri_2001,Oguri_2007}. 
Applying a magnetic field however causes the peak of the resonance to split, while the sidebands get farther apart due to the fact the singly occupied level is lowered in energy by $-H/2$ while the energy of the doubly occupied level increases by $H/2$. This agrees qualitatively with numerical and experimental results \cite{Anders_2008,Millis,Muehlbacher,Han_2010,Leturcq}.  As a final check of this method, our results reduce to the well-known Fermi liquid expansion at low bias, energies, temperature and magnetic field \cite{Nozieres_1974}.	

The scheme has a potential for being generalized to the non-perturbative regime using well-controlled techniques such as the large-N expansion \cite{Mitra,Read_Newns}, and such extensions are currently under investigation. In this case the Lippmann-Schwinger operators we use as our point of reference follow from an effectively non-interacting Hamiltonian, which arises from the large-$N$ saddle point solution of the functional integral obtained using the slave-boson formalism. There is scope to treat more sophisticated systems such as Luttinger liquids within this formalism by combining bosonization and scattering techniques. Also, a possible extension of the formalism to include thermoelectric transport in addition to bias-induced transport, is being explored. We emphasize that the effective
equilibrium approach is that it is conducive to the implementation of numerical methods, such as numerical RG \cite{Anders_2008,Anders_2010}
and Quantum Monte Carlo \cite{Han_2010}.

This method is independent of the exact nature of the interaction, provided it is localized in the vicinity of the quantum impurity, and can be used to treat other models such as the interacting resonant level. Furthermore, the extension to multiple leads and other more complicated geometries can be straightforwardly accomplished.

{\bf Acknowledgments:}
 We acknowledge N.\ Andrei, H.\ U.\ Baranger and T.\ L.\ Schmidt for stimulating discussions. This work is supported by the
Department of Energy under grant DE-FG02-08ER46541 (P.D. and K.L.H.), by the Yale Center for Quantum Information Physics (NSF DMR-0653377, J.K. and K.L.H.) and by the NSF under grant DMR-0907150 (J.E.H.).

\begin{appendix}

\section{The $Y$ operator}

In this Appendix we present an alternative derivation of the steady state density matrix. To build the $Y$-operator, the starting point is the expansion of the steady-state density matrix into a power series in the tunneling Hamiltonian $H_T$. This can be accomplished 
by employing the time-dependent framework of the open-system limit where the steady-state density matrix is obtained by adiabatically switching on the coupling in the far past. One obtains analytical expressions for the power series in $H_T$. Order for order comparison of these expansions then leads to a system of nested differential equations for the bias operators. Finally, the solution to this system of differential equations is then shown to be identical with the representation of $Y$ in terms of Lippmann-Schwinger operators, see Eq.\ \eqref{Yrep}. 

We should emphasize the fact that the expansion in powers of $H_{T}$ is a purely formal procedure, which we adopt for the sake of a systematic comparison, and the proof below is non-perturbative in $H_{T}$.

\subsection{Expansion of $\rho$ using the effective equilibrium representation}

In the derivation of the explicit form of the $Y$ operator, it is convenient to collect terms in orders of the tunneling Hamiltonian $H_{T}$. We thus start by expanding $Y=\sum_{n=0}^{\infty}Y_{n}$ into a series in powers of the tunneling, such that $Y_{n}\propto \left(H_{T}\right)^{n}$. In the following, the index $n$ will always be used for power counting of the tunneling Hamiltonian $H_{T}$. From Eq.\ \eqref{rhoHershfield} we thus obtain 
\begin{align}
\rho&=\exp\left[ -\beta\left\{(H_{0} - Y_{0})+(H_{T}-Y_{1})-\sum_{n=2}^{\infty}Y_{n}\right\}\right]\notag\\
&=\exp\left[-\beta \sum_{n=0}^{\infty}X_{n}\right].
\end{align}
Here, we have regrouped the Hamiltonian with the $Y$ operator order by order in the auxiliary operator $X_n\sim (H_T)^n$, which  is hence defined as
\be\label{XY}
X_n \equiv \begin{cases}
H_0 - Y_0, & n=0\\
H_T - Y_1, & n=1\\
- Y_n,     & n\ge2\ .
\end{cases}
\ee
 
One can expand the exponential operator to collect terms in powers of the tunneling $H_{T}$ such that 
\begin{align}
\rho&
=\sum_{l=0}^{\infty}\frac{(-\beta)^{l}}{l!}\sum_{i_{1},\ldots,i_{l}=0}^{\infty} X_{i_{1}}\ldots X_{i_{l}}\notag\\
&=\sum_{n=0}^{\infty}\sum_{l=0}^{\infty}\frac{(-\beta)^{l}}{l!}\sum_{i_{1}+\cdots+i_{l}=n} X_{i_{1}}\ldots X_{i_{l}} \equiv\sum_{n=0}^{\infty}\rho_{n},
\label{rho-exp}
\end{align}
where, following our general notation, $\rho_n$ denotes the order $(H_T)^n$ contribution to the steady-state density matrix.

\subsection{Expansion of $\rho$ using the open-system approach}

Let us now employ the open-system approach (as outlined in Section 2). In this case, the steady-state density matrix  is obtained by switching on the tunneling in the early past $t_0<0$. Up to this time $t_0$, the leads are decoupled from the dot. The adiabatic switch-on of tunneling is facilitated by using $H_{T}e^{\eta t}\theta(t-t_{0})$ for the tunneling Hamiltonian (note that in this scenario, the total Hamiltonian is therefore time dependent). At time $t=0$, transients have decayed and time evolution has turned the original density matrix $\rho_0=\bar{\rho}(t=t_0)$ into the steady-state density matrix $\rho=\bar{\rho}(t=0)$.  We note that the condition of adiabaticity is strictly true only in the limit $\frac{1}{\vert t_{0}\vert}\ll\eta$, where $\vert t_{0}\vert\rightarrow \infty$, which is assumed in the open-system limit \cite{Mehta_Andrei_new}. We emphasize that the actual evaluation of this limit is deferred until the very end of all calculations. Keeping $1/|t_{0}|$ and $\eta$ small but nonzero in the interim is crucial for mathematical clarity and for avoiding ill-defined expressions.

This time, it will be convenient to work in the interaction picture (with respect to the tunneling), where
in general 
\be
\mathcal{O}_{I}(t)=e^{iH_{0}(t-t_{0})}\mathcal{O}e^{-iH_{0}(t-t_{0})}.
\ee
denotes the interaction picture of the operator $\mathcal{O}$. Note that the time $t_0$ (not $t=0$) has been chosen as the reference time where Schr\"odinger and interaction pictures agree, $\mathcal{O}=\mathcal{O}_I(t_0)$. [In this we differ from the conventions adopted by Hershfield \cite{hershfield_reformulation_1993} who chose different reference times for the Heisenberg and interaction representation.] 

In the interaction picture, the density matrix satisfies the evolution equation
\begin{equation}
i\frac{d}{dt}\bar{\rho}_{I}(t)=[H_{T,I}(t),\bar{\rho}_{I}(t)].
\label{eq:densitymatrix}
\end{equation}
This is formally solved by
\begin{align}\label{eq:interactionsolution}
\bar{\rho}_{I}(t)&=\sum_{n=0}^{\infty}\frac{(-i)^n}{n!}\mathcal{T}\bigg\{\prod_{i=1}^{n}\left[\int_{t_{0}}^{t}dt_{i}\right][H_{T,I}(t_{1}),[H_{T,I}(t_{2}),[\ldots[H_{T,I}(t_{n}),\rho_{0}]]\ldots]\bigg\}\notag\\
&\equiv \sum_{n=0}^\infty \bar{\rho}_{n,I}(t),
\end{align}
where the $n=0$ term simply fixes the boundary condition $\bar{\rho}(t=t_0)=\rho_0$ and $\mathcal{T}$ denotes time-ordering of operators.

It should be noted that Eq.\ \eqref{eq:interactionsolution} already has the form of a power series in the tunneling and thus has been used to define the $n$-th order contribution $\bar{\rho}_{n,I}(t)$ to the interaction-picture density matrix. For later purposes it is useful to note that the contributions can alternatively be obtained from
\begin{align}
\frac{d}{dt}\bar{\rho}_{n,I}(t)=-i[H_{T,I}(t),\bar{\rho}_{n-1,I}(t)],
\label{eq:nested}
\end{align}
i.e., a system of nested differential equations associated with the boundary conditions $\bar{\rho}_{0,I}(t_0)=\rho_0$ and $\bar{\rho}_{n,I}(t_0)=0$ for $n\ge1$.

\subsection{Derivation of differential equations for $Y_n$}

For the results to be consistent, we require (assuming steady state is reached \cite{Doyon_Andrei}) that the steady-state density matrix $\rho$ be identical to the steady-state density matrix obtained in the open-system limit. This allows one to determine the correct form of the $Y$ operator, order for order in the tunneling. 

We thus impose the identity of the steady-state density matrices
\begin{equation}
e^{iH_{0}(t-t_{0})}\rho_{n}e^{-iH_{0}(t-t_{0})}=\bar{\rho}_{n,I}(t), 
\label{eq:interactionpic}
\end{equation}
where we have transformed $\rho_n$ into the interaction picture. Eq.\ \eqref{eq:interactionpic} is expected to hold for all $t>0$, since at that point the interaction has been fully switched on, and the time-independent Hamiltonian (used in Hershfield's effective equilibrium approach) and the time-dependent Hamiltonian (from the open-system approach) are identical.

We now differentiate the left-hand side of Eq.\ \eqref{eq:interactionpic} with respect to time and use Eq.\ \eqref{rho-exp} to obtain
\begin{align}
\label{eq:diff1}
&\frac{d}{dt}\rho_{n,I}(t)=\sum_{l=1}^{\infty}\frac{(-\beta)^{l}}{l!}\sum_{i_{1}+\cdots+i_{l}=n}\; 
\sum_{k=1}^{l} X_{i_{1},I}\cdots\frac{dX_{i_{k},I}(t)}{dt}\cdots X_{i_{l},I},
\end{align}
where $X_{i_{k},I}=e^{iH_{0}(t-t_{0})}X_{i_{k}}e^{-iH_{0}(t-t_{0})}$ denotes the interaction picture representation of $X_{i_{k}}$.
Similarly, we may differentiate the right-hand side of Eq.\ \eqref{eq:interactionpic}. The resulting commutator is given in Eq. \eqref{eq:nested} and contains $\rho_{n-1,I}(t)$,\footnote{Note that due to the identity \eqref{eq:interactionpic} we may, from here on, drop all bars on $\rho$.} for which we substitute the corresponding expression from Eq.\ \eqref{rho-exp}. This way, we obtain

\begin{align}
\frac{d}{dt} \rho_{n,I}(t)&=i[\rho_{n-1,I}(t),H_{T,I}(t)]\notag\\
&=i\sum_{l=1}^{\infty}\frac{(-\beta)^{l}}{l!}\sum_{i_{1}+i_{2}+\ldots+i_{l}=n-1}
\;\sum_{k=1}^{l} X_{i_{1},I}(t)\ldots[X_{i_{k},I}(t),H_{T,I}(t)]\ldots X_{i_{l},I}(t)\notag\\\label{eq:diff2}
&=i\sum_{l=1}^{\infty}\frac{(-\beta)^{l}}{l!}\sum_{i_{1}+i_{2}+\ldots+i_{l}=n}
\;\sum_{k=1}^{l}X_{i_{1},I}(t)\ldots[X_{i_{k}-1,I}(t),H_{T,I}(t)]\ldots X_{i_{l},I}(t),
\end{align}
where for $n<0$ we define $X_n=0$.
In the last step of Eq.\ \eqref{eq:diff2}, the summation constraint is shifted from $n-1$ to $n$ to facilitate the comparison with Eq.\ \eqref{eq:diff1}. This comparison yields the relation
\begin{align}
\frac{d}{dt}X_{n,I}=i[X_{n-1,I}(t),H_{T,I}(t)].
\end{align}
Finally, utilizing the relation \eqref{XY} between $X$ and $Y$ operators, one finds that the $Y_{n}$ operators also satisfy
\begin{align}
\frac{d}{dt}Y_{n,I}=i[Y_{n-1,I}(t),H_{T,I}(t)].
\label{eq:Yoperatordiff}
\end{align}
To obtain the $Y$ operator from these differential equations, it is crucial to specify the boundary conditions at the initial time $t=t_0$. Given the relation $e^{\mathcal{O}_{I}(t)}=e^{iH_{0}(t-t_{0})}e^\mathcal{O} e^{-iH_{0}(t-t_{0})}$, valid for any operator $\mathcal{ O}$, the boundary condition $\bar{\rho}(t=t_{0})=\rho_{0}$ implies \cite{Doyon_Andrei}
\be\label{boundcond}
\lim_{t\searrow t_{0}}Y_{I}(t)=\frac{\Phi}{2}\sum_{\alpha}\alpha N_{\alpha}.
\ee
It now remains to prove that the following interaction-picture expression of $Y$
 \begin{align}
 Y_{I}(t)=&\frac{\Phi}{2}\sum_{\alpha k \sigma}\alpha \psi^{\dag}_{\alpha k\sigma,I}(t)\psi_{\alpha k\sigma,I}(t),
\label{eq:Y operator}
\end{align}
in terms of the Lippmann-Schwinger operators $\psi^{\dag}_{\alpha k\sigma,I}(t)$ represents the solution to the above initial-value problem. Once we have recapitulated the crucial properties of these Lippmann-Schwinger operators in the following subsection, it will be simple to confirm that this ansatz indeed solves the differential equation \eqref{eq:Yoperatordiff} subject to the boundary condition \eqref{boundcond}.

\subsection{Properties of the Lippmann-Schwinger operators}

It has been shown \cite{Han_2006} for the generic model defined in Section 2 that the Lippmann-Schwinger operators are fermionic,
\begin{align}
\left\{\psi^\dagger_{\alpha k\sigma},\psi_{\alpha'k'\sigma'}\right\}=\delta_{\alpha\alpha'}\delta_{kk'}\delta_{\sigma\sigma'},
\label{eq:LSoperator}
\end{align}
and diagonalize the full Hamiltonian (including all interactions),
\begin{align}
H=\sum_{\alpha k}\epsilon_{\alpha k}\psi^{\dag}_{\alpha k\sigma}\psi_{\alpha k\sigma}.
\end{align}
One can expand $\psi^{\dag}_{\alpha k\sigma}$ in powers of the tunneling Hamiltonian $H_{T}$, {\it i.e.}, 
\be
\psi^{\dag}_{\alpha k\sigma}=\sum_{n=0}^{\infty}\psi^{\dag}_{\alpha k\sigma,n}
\ee
such that $\psi^{\dag}_{\alpha k\sigma,n}\propto \left(H_{T}\right)^{n}$. In the interaction representation, the operators $\psi^{\dag}_{\alpha k\sigma}$ satisfy
\begin{align}
\frac{d}{dt}\psi^{\dag}_{\alpha k\sigma,I}(t)&=i[H_{0},\psi^{\dag}_{\alpha k\sigma,I}(t)]=i[H_{I}(t),\psi^{\dag}_{\alpha k\sigma,I}(t)]-i[H_{T,I}(t),\psi^{\dag}_{\alpha k\sigma,I}(t)],
\label{eq:diffeqnth}
\end{align}
subject to the boundary condition 
\be
\lim_{t\searrow t_0} \psi_{\alpha k \sigma,I}(t) = c_{\alpha k \sigma}.
\ee
Eq. \eqref{eq:diffeqnth} can be further simplified, and collecting orders of $(H_{T})^{n}$, cast into the set of nested differential equations
\be
\frac{d}{dt}\psi^{\dag}_{\alpha k\sigma,n,I}(t)=i\epsilon_{\alpha k}\psi^{\dag}_{\alpha k\sigma,n,I}(t)+i[\psi^{\dag}_{\alpha k\sigma,n-1,I}(t),H_{T,I}(t)]. 
\ee

As discussed in Ref.\ \cite{Han_2006}, the formal solution can be expressed compactly as
\begin{equation}
\psi^\dagger_{\alpha k\sigma}=c^\dagger_{\alpha k\sigma}
+\frac{1}{\epsilon_{\alpha k}-\mathcal{ L}+i\eta}\mathcal{ L}_{T} 
c^\dagger_{\alpha k\sigma}.
\end{equation}
From Eq.\ \eqref{eq:lippmannformal} one can write the detailed form of the operator $\psi^{\dag}_{\alpha k\sigma,n}$
\begin{align}
\psi^\dagger_{\alpha k\sigma}=\sum_{n=0}^{\infty}\left[\left(\frac{1}{\epsilon_{\alpha k}-\mathcal{ L}_{0}+i\eta}\mathcal{ L}_{T}\right)^{n}c^\dagger_{\alpha k\sigma}\right]\equiv\sum_{n}\psi^\dagger_{\alpha k\sigma,n}.
\end{align}
Note that when the tunneling is set to zero, one finds indeed that $\psi^\dagger_{\alpha k\sigma}\rightarrow c^\dagger_{\alpha k\sigma}$.

\subsection{Conclusion of the proof}

It is now simple to verify that the ansatz for $Y_{n,I}(t)$ solves the set of differential equations given by Eq.\ (\ref{eq:Y operator}), and obeys the appropriate boundary condition \eqref{boundcond}: We start by writing the $(H_T)^n$ contribution to the $Y$ operator as
\begin{align}
Y_{n,I}(t)=\sum_{\alpha k\sigma}\alpha\frac{\Phi}{2}\sum_{p=0}^{n}\psi^{\dag}_{\alpha k\sigma,p,I}(t)\psi_{\alpha k\sigma,n-p,I}(t).
\end{align}
Differentiating this with respect to time yields
\begin{align}
\frac{d}{dt}Y_{n,I}(t)&=\sum_{\alpha k\sigma}\alpha\frac{\Phi}{2}\sum_{p=0}^{n}\bigg[\left(\frac{d}{dt}\psi^{\dag}_{\alpha k\sigma,p,I}(t)\right)\psi_{\alpha k\sigma,n-p,I}(t)\notag\\
&\qquad\qquad\qquad\qquad+\psi^{\dag}_{\alpha k\sigma,p,I}(t)\left(\frac{d}{dt}\psi_{\alpha k\sigma,n-p,I}(t)\right)\bigg],
\end{align}
so that in conjunction with Eq.\ (\ref{eq:diffeqnth}) one obtains
\begin{align}
&\frac{d}{dt}Y_{n,I}(t)=\sum_{\alpha k\sigma}\alpha\frac{\Phi}{2}\sum_{p=0}^{n-1}[\psi^{\dag}_{\alpha k\sigma,p,I}(t)\psi_{\alpha k\sigma,n-1-p,I}(t)]=i[Y_{n-1,I}(t),H_{t,I}].
\end{align}
This concludes the proof.

In summary, we find that the steady state dynamics of the system can be described by an effective equilibrium density matrix of the form
\begin{align}
{\rho}&=e^{-\beta(H - Y)},
\end{align}
where the operator
\begin{align}
Y=\frac{\Phi}{2}\sum_{\alpha k \sigma}\alpha\psi^{\dag}_{\alpha k\sigma}\psi_{\alpha k\sigma}
\end{align}
encodes the entire nonequilibrium boundary condition of the system. 

\subsection{Recursion relations for expectation values of observables}

In addition to the previous proof, we follow Hershfield \cite{hershfield_reformulation_1993} and underpin the equivalence of the adiabatic approach and the effective equilibrium approach by showing that they lead to identical recursion relations for expectation values of observables. Again, the systematic use of the open-system limit makes the proof sound.

Let $\mathcal{ O}$ denote a generic observable. To derive the first recursion relation, we will decompose the steady-state expectation value $\langle\mathcal{ O}\rangle$ into a series counting the powers of the tunneling Hamiltonian $H_T$. In the first step, we thus substitute the expansion \eqref{rho-exp} of $\rho$,
\begin{align}
\langle\mathcal{ O}\rangle&=\frac{\Tr[\rho\mathcal{ O}]}{\Tr[\rho]}
=\frac{\Tr\left[\sum_{n=0}^{\infty}\rho_{n}\mathcal{ O}\right]}{\Tr\left[\sum_{m=0}^{\infty}\rho_{m}\right]}.
\end{align}
Pulling out a factor of $1/\Tr[\rho_0]$, expanding the denominator as a geometric series, and collecting terms order for order in $H_T$, we can rewrite this as
\begin{align}
\langle\mathcal{ O}\rangle=&\sum_{n=0}^{\infty} \sum_{l=0}^{n}(-1)^{l}\sideset{}{'}\sum_{j_{1},\ldots,j_{l}=1}^\infty \prod_{s=1}^l
\left[\frac{\Tr[\rho_{j_{s}}]}{\Tr[\rho_{0}]}\right]
\frac{\Tr[\rho_{n-\sum_{s=1}^l j_{s}}\mathcal{ O}]}{\Tr[\rho_{0}]}\nonumber\\
\equiv&\sum_{n=0}^{\infty}\langle\mathcal{O}\rangle_{n},\label{on-eq}
\end{align}
where $\sum'$ denotes a restricted summation, subject to the condition $\sum_{s=1}^l j_{s}\le n$. Eq. \eqref{on-eq} allows one to prove the important recursion relation
\begin{align}
\langle\mathcal{ O}\rangle_{n}=\frac{\Tr[\rho_{n}\mathcal{ O}]}{\Tr[\rho_{0}]}-\sum_{k=1}^{n}\frac{\Tr[\rho_{k}]}{\Tr[\rho_{0}]}\langle\mathcal{ O}\rangle_{n-k},
\label{eq:timeindep}
\end{align}
which relates the $n$-th order term to the expectation value of $\mathcal{O}$ with respect to $\rho_n$, and all lower-order terms $\langle\mathcal{ O}\rangle_{m}$ ($m=0,\ldots,n-1$).

Now, we turn to the time-dependent representation assuming an adiabatic switch-on of the tunneling.
The steady-state expectation value of an observable may now be obtained via
\begin{equation}
\langle \mathcal{ O}\rangle=\frac{\Tr[\bar{\rho}_{I}(0)\mathcal{ O}_{I}(0)]}{\Tr[\rho_{0}]},\label{oi-exp}
\end{equation}
where we have switched to the interaction picture.
To arrive at the desired recursion relation, we again decompose this into a power series of the tunneling Hamiltonian,
\be
\langle \mathcal{ O}\rangle=\frac{\Tr[\bar{\rho}_{I}(0)\mathcal{ O}_{I}(0)]}{\Tr[\rho_{0}]} = \sum_{n=0}^\infty \langle \mathcal{ O}\rangle_n,
\ee 
where
\begin{align}\label{t-on}
&\langle\mathcal{ O}\rangle_{n}=\frac{(-i)^n}{n!}\frac{1}{\Tr\rho_{0}}\Tr\bigg\{\mathcal{ T}\prod_{i=1}^{n}\left[\int_{t_{0}}^{t=0}dt_{i}\right]\\\notag
&\quad\times[H_{T,I}(t_{1}),[H_{T,I}(t_{2}),[\ldots[H_{T,I}(t_{n}),\rho_{0}]]\ldots]\mathcal{ O}_{I}(0)\bigg\}.
\end{align}	
This expression for $\langle\mathcal{ O}\rangle_{n}$ allows one to derive the second recursion relation, which reads
\begin{align}
&\langle \mathcal{ O}\rangle_{n}=\frac{\Tr[\rho_{n,I}(0)\mathcal{ O}_{I}(0)]}{\Tr[\rho_{0}]} -\sum_{k=1}^{n}\frac{\Tr[\rho_{k,I}(0)]}{\Tr[\rho_{0}]}\langle\mathcal{ O}\rangle_{n-k}.
\label{eq:timedep}
\end{align}
The details of this derivation are given in Appendix \ref{app:1}. The proof makes explicit use of the factorization of two-time correlation functions in the limit of large time separation \cite{Doyon_Andrei}. The agreement between Eqs. (\ref{eq:timeindep}) and (\ref{eq:timedep}) underpins the equivalence between the time-dependent adiabatic approach and the effective equilibrium approach.

\section{Derivation of the recursive form of $\mathcal{ O}_{n}$}
\label{app:1}

In this Appendix we provide the detailed derivation of the recursion relation \eqref{eq:timedep} when working within the time-dependent approach, switching on the tunneling adiabatically. Our starting point for the derivation is the expression for the order-$(H_T^n)$ contribution to the steady-state expectation value of some operator $\mathcal{O}$,
\begin{align}
\langle \mathcal{ O}\rangle_{n}=&\frac{(-i)^{n}}{\Tr[\rho_{0}]}\Tr\bigg\{\int_{t_{0}}^{0}dt_{n}\int_{t_{0}}^{t_{n}}dt_{n-1}\cdots\int_{t_{0}}^{t_{2}}dt_{1}\notag\\
[H_{T,I}(t_{n})&,[H_{T,I}(t_{n-1}),[\ldots[H_{T,I}(t_{1}),\rho_{0}]]\ldots]\mathcal{ O}_{I}(t=0)\bigg\},
\label{eq:nthterm}
\end{align}
see Eq.\ \eqref{t-on} in the main text. Carrying out the $t_{1}$ integration and using Eq.\ (\ref{eq:densitymatrix}), one obtains
\begin{align}
&\langle \mathcal{ O}\rangle_{n}=\frac{(-i)^{n-1}}{\Tr[\rho_{0}]}\Tr\bigg[\int_{t_{0}}^{0}dt_{n}\int_{t_{0}}^{t_{n}}dt_{n-1}\cdots\int_{t_{0}}^{t_{3}}dt_{2}\\
&\times[H_{T,I}(t_{n}),[\ldots[H_{T,I}(t_{2}),\left(\rho_{1,I}(t_{2})-\rho_{1,I}(t_{0})\right)]\ldots]\mathcal{ O}_{I}(0)\bigg].\notag
\end{align}
We now retain the term involving $\rho_{1,I}(t_{0})$, and proceed by carrying out the $t_{2}$ integration for the term containing $\rho_{1,I}(t_{2})$. This integration yields 2 terms: one with $\rho_{2,I}(t_{0})$ which we retain, and the other with $\rho_{2,I}(t_{3})$ which we subject to further integration. Continuing in this fashion until all the variables $t_{1},\ldots,t_{n}$ in the latter term have been integrated out, we arrive at the result 
\begin{align}
&\langle\mathcal{ O}\rangle_{n}=\frac{\Tr\left[\rho_{n,I}(0)\mathcal{ O}_{I}(0)\right]}{\Tr[\rho_{0}]}\\\notag
&-\sum_{p=1}^{n}\frac{(-i)^{n-p}}{\Tr[\rho_{0}]}\int_{t_{0}}^{t_{n+1}=0}dt_{n}\int_{t_{0}}^{t_{n}}dt_{n-1}\cdots\int_{t_{0}}^{t_{p+2}}dt_{p+1}\notag\\
&\times\Tr\bigg\{[H_{T,I}(t_{n}),[\ldots,[H_{T,I}(t_{p+1}),\rho_{p,I}(t_{0})]\ldots]\mathcal{ O}_{I}(0)\bigg\}.
\end{align}
Let us focus on the $p$-th term of the above sum. One observes that when $p=n$, none of the integrals are present and the argument of the trace is just $\rho_{p,I}(t_{0})\mathcal{ O}_{I}(0)$. Suppressing the integrals and other c-numbers we isolate the product of operators within the trace 
\begin{align}
&\Tr\bigg\{[H_{T,I}(t_{n}),[\ldots,[H_{T,I}(t_{p+1}),\rho_{p,I}(t_{0})]]\ldots]\mathcal{ O}_{I}(0)\bigg\}\notag.\\
\intertext{We can cycle $\rho_{p,I}(t_{0})$ in the above expression out of the nested commutators, thereby inverting the nested commutator structure to re-express this as}
&\Tr\bigg\{\rho_{p,I}(t_{0})[\ldots[\mathcal{ O}_{I}(0),H_{T,I}(t_{n})],...H_{T,I}(t_{p+1})]\bigg\}.
\end{align}
This allows us to rewrite $\langle\mathcal{O}\rangle_n$ in the form
\begin{align}\label{eq:form1}
&\langle\mathcal{ O}\rangle_{n}=\frac{\Tr\left[\rho_{n,I}(0)\mathcal{ O}_{I}(0)\right]}{\Tr[\rho_{0}]}\\\notag
&-\sum_{p=1}^{n}\frac{(-i)^{n-p}}{\Tr[\rho_{0}]}\int_{t_{0}}^{0}dt_{n}\int_{t_{0}}^{t_{n}}dt_{n-1}\cdots\int_{t_{0}}^{t_{p+2}}dt_{p+1}\\\notag
&\times\Tr\bigg[\rho_{p,I}(t_{0})[\ldots[\mathcal{ O}_{I}(0),H_{T,I}(t_{n})],\ldots,H_{T,I}(t_{p+1})]\bigg].
\end{align}
To proceed further, let
\begin{align}
\langle\langle\cdots\rangle\rangle_{0}=\frac{\Tr\left(e^{-\beta(H_{0}-Y_{0})}\cdots\right)}{\Tr\left(e^{-\beta(H_{0}-Y_{0})}\right)},
\end{align}
denote an expectation value with respect to the density matrix $\rho_{0}$. It has been shown by Doyon and Andrei in Ref.\ \cite{Doyon_Andrei} that, in the open system limit, long-time correlation functions factorize (this argument is rigorous at finite temperatures), \emph{i.e.}
\begin{equation}
\langle\langle\mathcal{ O}(t_{1})\mathcal{ O}(t_{2})\rangle\rangle_{0}\rightarrow\langle\langle\mathcal{ O}(t_{1})\rangle\rangle_{0}\langle\langle\mathcal{ O}(t_{2})\rangle\rangle_{0}
\label{eq:factorize}
\end{equation}
as $\vert t_{1}-t_{2}\vert\rightarrow\infty$. We will now employ this factorization in Eq.\ (\ref{eq:form1}). Renaming the integration variables $t'_{1}\equiv t_{p+1}$, $t'_{2}\equiv t_{p+2}$, $\ldots$, $t'_{n-p}\equiv t_{n}$ and inserting $\rho_{0}\rho_{0}^{-1}=\mathbf{1}$ we get

\begin{align}
&\langle\mathcal{ O}\rangle_{n}=\frac{\Tr\left[\rho_{n,I}(0)\mathcal{ O}_{I}(0)\right]}{\Tr[\rho_{0}]}\\\notag
&-\sum_{p=1}^{n}\frac{(-i)^{n-p}}{\Tr[\rho_{0}]}\int_{t_{0}}^{0}dt'_{n-p}\int_{t_{0}}^{t'_{n-p}}dt'_{n-p-1}\cdots\int_{t_{0}}^{t'_{2}}dt'_{1}
\Tr\bigg\{\rho_{0}\underbrace{\rho_{0}^{-1}\rho_{p,I}(t_{0})}_{A(t_{0})}\notag\\
&\qquad\qquad\qquad\times\underbrace{[[\ldots[\mathcal{ O}_{I}(0),H_{T,I}(t'_{n-p})],\ldots],H_{T,I}(t'_{1})]}_{B(t'_{1},\ldots,t'_{n-p},t=0)}\bigg\}.
\label{eq:factor1}
\end{align}

We now recall that in the open system limit we take $v_{F}/L\ll1/\vert t_{0}\vert\ll\eta$ with $t_{0}\rightarrow-\infty$. This implies that the $e^{\eta t}$ term in $H_{T,I}$ essentially cuts off the lower limit of the time integrals at a time $\sim1/\eta$ much earlier than $t_{0}$. Thus, Eq.\ \eqref{eq:factor1} can be cast into an integral over $\langle\langle A(t_{0})B(t'_{1},\ldots,t'_{n-p},t=0)\rangle\rangle_{0}$, where $\vert t'_{1}-t_{0}\vert$,$\ldots$,$\vert t'_{n-p}-t_{0}\vert$,$\vert 0-t_{0}\vert\rightarrow\infty$. Using Eq.\ (\ref{eq:factorize}) we get
\begin{align}
&\langle\langle A(t_{0})B(t'_{1},\ldots,t'_{n-p},t=0)\rangle\rangle_{0}=\langle\langle A(t_{0})\rangle\rangle_{0}\langle\langle B(t'_{1},\ldots,t'_{n-p},t=0)\rangle\rangle_{0},
\end{align}
and hence
\begin{align}
&\langle\mathcal{ O}\rangle_{n}=\frac{\Tr\left[\rho_{n,I}(0)\mathcal{ O}_{I}(0)\right]}{\Tr[\rho_{0}]}\\\notag
&-\sum_{p=1}^{n}\frac{\Tr[\rho_{p,I}(t_{0})]}{\Tr[\rho_{0}]}\Tr\bigg[\big(-i)^{n-p}\int_{t_{0}}^{0}dt'_{n-p}\int_{t_{0}}^{t'_{n-p}}dt'_{n-p-1}
\cdots\int_{t_{0}}^{t'_{2}}dt'_{1}\notag\\
&\qquad\qquad\qquad[[\ldots[\mathcal{ O}_{I}(0),H_{T,I}(t'_{n-p})]\ldots,]H_{T,I}(t'_{1})]{\rho}_{0}\bigg].\notag
\end{align}
Inverting the nested commutator structure as previously, we get
\begin{align}
\langle\mathcal{ O}\rangle_{n}&=\frac{\Tr\left[\rho_{n,I}(0)\mathcal{ O}_{I}(0)\right]}{\Tr[\rho_{0}]}\\\notag
&-\sum_{p=1}^{n}\frac{\Tr[\rho_{p,I}(t_{0})]}{\Tr[\rho_{0}]}\Tr\bigg[\big(-i)^{n-p}\int_{t_{0}}^{0}dt'_{n-p}\int_{t_{0}}^{t'_{n-p}}dt'_{n-p-1}
\cdots\int_{t_{0}}^{t'_{2}}dt'_{1}\notag\\
&\qquad\qquad[H_{T,I}(t'_{n-p}),[\ldots,[H_{T,I}(t'_{1}),{\rho}_{0}]]\ldots]\mathcal{ O}_{I}(0)\bigg].
\end{align} 
Finally, comparing the second term in the latter equation with Eq.\ (\ref{eq:nthterm}) and noting that $\Tr[\rho_{p,I}(t_{0})]=\Tr[\rho_{p,I}(0)]$, we arrive at the final recursion relation
\begin{align}
\langle\mathcal{ O}\rangle_{n}=\frac{\Tr\left[\rho_{n,I}(0)\mathcal{ O}_{I}(0)\right]}{\Tr[\rho_{0}]}-\sum_{p=1}^{n}\frac{\Tr[\rho_{p,I}(0)]}{\Tr[\rho_{0}]}\langle\mathcal{ O}\rangle_{n-p}.
\end{align}

\section{Two derivations of $n_{d}$}

\subsection{Derivation of $n_{d}$ using the spectral representation}
\label{app:2}

Here, we provide additional details on the equivalence between the effective equilibrium approach and the Schwinger-Keldysh formalism and explicitly show the identity of the electron occupancy on the quantum dot as obtained in the Schwinger-Keldysh formalism and as obtained with the effective equilibrium approach. 

We start by stating two useful identities involving Liouvillian superoperators.
The first identity regards the imaginary-time Heisenberg representation of an operator $\mathcal{ O}$ and is given by
\begin{align}
 \mathcal{ O(\tau)}=e^{\tau(H-Y)}\mathcal{ O}e^{-\tau(H-Y)}=e^{\tau(\mathcal{ L}-\mathcal{ L}_{Y})}\mathcal{O}.
\end{align}
The second identity is a small trick which transfers a Liouvillian superoperator from one side of an anticommutator to the other,
\begin{align}
\left\langle \left\{  \mathcal{ O}_{1}, \frac{1}{z-\mathcal{ L}}\mathcal{ O}_{2} \right\} \right\rangle
=\left\langle  \left\{  \frac{1}{z+\mathcal{ L}}\mathcal{ O}_{1}, \mathcal{ O}_{2} \right\} \right\rangle,
\label{appb-eq}
\end{align}
where $z$ is a c-number.

We now set out to demonstrate the equivalence of the Schwinger-Keldysh expression for the dot occupancy,
\begin{align}\label{nd-keldysh}
n_{d}&=\frac{1}{2}\sum_{\sigma}\int_{-\infty}^{\infty}d\epsilon\,  A_{d_{\sigma}}(\epsilon)
\left[f\left(\epsilon+\frac{\Phi}{2}\right)+f\left(\epsilon-\frac{\Phi}{2}\right)\right]\nonumber\\
&=\frac{1}{2\nu(0)\Omega}\sum_{\sigma k}  A_{d_{\sigma}}(\epsilon_k)
\left[f\left(\epsilon_k+\frac{\Phi}{2}\right)+f\left(\epsilon_k-\frac{\Phi}{2}\right)\right]
\end{align}
and the expression obtained in the effective equilibrium approach,
\be \label{nd-effeq}
n_d = \sum_\sigma \langle d_\sigma^\dag d_\sigma \rangle =\sum_{\sigma}\mathcal{ G}_{d_{\sigma}{d}_{\sigma}^\dag}(\tau=0).
\ee
It is crucial to recall the definition of the Green's functions in the effective equilibrium approach (see Section \ref{sec:correspondence}). The Fourier representation of the (imaginary) time ordered Green's function is given by
\begin{align}
\mathcal{ G}_{\mathcal{ O}_{1}\mathcal{ O}_{2}}(i\omega_{n})=\left\langle\left\{\mathcal{ O}_{1},\frac{e^{i\omega_{n}0^{{+}}}}{i\omega_{n}-\mathcal{ L}+\mathcal{ L}_{Y}}\mathcal{ O}_{2}\right\}\right\rangle,
\end{align}
and the retarded/advanced real-time Green's function in frequency space is obtained by
\begin{align}\label{g-retadv}
G^{\text{ret/adv}}_{\mathcal{ O}_{1}\mathcal{ O}_{2}}(\omega)&=\left\langle\left\{\mathcal{ O}_{1},\frac{1}{\omega\mp\mathcal{ L}\pm i\eta}\mathcal{ O}_{2}\right\}\right\rangle.
\end{align}

To prove the equivalence between Eqs.\ \eqref{nd-keldysh} and \eqref{nd-effeq}, we start with the Keldysh expression and evaluate the spectral function 
\be
A_{d_{\sigma}}(\epsilon)=-\frac{1}{\pi}\iim G^{\text{ret}}_{d_{\sigma}d_{\sigma}^\dag}(\epsilon) = \frac{1}{2\pi i}\bigg[ G^{\text{adv}}_{d_{\sigma}d_{\sigma}^\dag}(\epsilon)-G^{\text{ret}}_{d_{\sigma}d_{\sigma}^\dag}(\epsilon)\bigg].
\ee
With Eq.\ \eqref{g-retadv} we evaluate the retarded and advanced Green's functions involved, 
\begin{align}
G^{\text{ret}}_{d_{\sigma}d_{\sigma}^\dag}(\epsilon_k)
= \left\langle\left\{d
_{\sigma},\frac{1}{\epsilon_k-\mathcal{ L}+i\eta}d^{\dag}_{\sigma}\right\}\right\rangle
= \frac{\sqrt{\Omega}}{t}
\langle\{d_{\sigma},\psi^{\dag}_{\alpha k\sigma}\}\rangle,
\end{align}
where, for simplicity, we are considering symmetric tunnel couplings and have set $t_1=t_{-1}=t$.
Similarly, one finds for the advanced Green's function
\begin{align}
G^{\text{adv}}_{d_{\sigma}d_{\sigma}^\dag}(\epsilon_k)
=\left\langle\left\{d
^{\dag}_{\sigma},\frac{1}{\epsilon_k-\mathcal{ L}+i\eta}d_{\sigma}\right\}\right\rangle
=\frac{\sqrt{\Omega}}{t}\langle\{d^{\dag}_{\sigma},\psi_{\alpha k\sigma}\}\rangle. 
\end{align}
Utilizing these relations in the evaluation of the occupancy, we find
\begin{align}
&n_{d}=\frac{1}{4\pi i}\frac{1}{t\nu(0)\sqrt{\Omega}}\sum_{\alpha k \sigma} \left(\langle\{d^{\dag}_{\sigma},\psi_{\alpha k\sigma}\}\rangle-\text{h.c.}\right)
f\left(\epsilon_k-\alpha\frac{\Phi}{2}\right).
\end{align}
This is equivalent to:
\begin{align}
&n_{d}=\frac{1}{4\pi i}\frac{1}{t\nu(0)\sqrt{\Omega}}\frac{1}{\beta}\sum_{\alpha k \sigma \omega_{n}} \frac{e^{i\omega_{n}0^{{+}}}}{i\omega_{n}-\epsilon_k+\alpha\frac{\Phi}{2}}
\left(\langle\{d^{\dag}_{\sigma},\psi_{\alpha k\sigma}\}\rangle-\langle\{d_{\sigma},\psi^{\dag}_{\alpha k\sigma}\}\rangle\right)\notag\\
&=\frac{1}{4\pi i t\nu(0)\sqrt{\Omega}}\frac{1}{\beta}\sum_{\alpha k \sigma \omega_{n}} \bigg(\left\langle\left\{d^{\dag}_{\sigma},\frac{e^{i\omega_{n}0^{{+}}}}{i\omega_{n}+\mathcal{ L}-\mathcal{ L}_{Y}}\psi_{\alpha k\sigma}\right\}\right\rangle 
-\left\langle\left\{\frac{e^{i\omega_{n}0^{{+}}}}{i\omega_{n}-\mathcal{ L}+\mathcal{ L}_{Y}}\psi^{\dag}_{\alpha k\sigma},d_{\sigma}\right\}\right\rangle\bigg)\notag\\
&=\frac{1}{4\pi i t\nu(0)\sqrt{\Omega}}\frac{1}{\beta}\sum_{\alpha k \sigma \omega_{n}} \bigg(\left\langle\left\{d^{\dag}_{\sigma},\frac{e^{i\omega_{n}0^{{+}}}}{i\omega_{n}+\mathcal{ L}-\mathcal{ L}_{Y}}c_{\alpha k\sigma}\right\}\right\rangle  \notag\\
&+ \left\langle\left\{d^{\dag}_{\sigma},\frac{e^{i\omega_{n}0^{{+}}}}{i\omega_{n}+\mathcal{ L}-\mathcal{ L}_{Y}}\frac{t}{\sqrt{\Omega}}\frac{1}{\epsilon_k+\mathcal{ L}-i\eta}d_{\sigma}\right\}\right\rangle\notag\\
& -\left\langle\left\{\frac{e^{i\omega_{n}0^{{+}}}}{i\omega_{n}-\mathcal{ L}+\mathcal{ L}_{Y}}c^{\dag}_{\alpha k\sigma},d_{\sigma}\right\}\right\rangle
-\left\langle\left\{\frac{e^{i\omega_{n}0^{{+}}}}{i\omega_{n}-\mathcal{ L}+\mathcal{ L}_{Y}}\frac{t}{\sqrt{\Omega}}\frac{1}{\epsilon_k-\mathcal{ L}+i\eta}d^{\dag}_{\sigma},d_{\sigma}\right\}\right\rangle\bigg).
\end{align}

The first and third terms can be combined, and recalling Eq.\ \eqref{eq:current1} they can be shown to cancel,
\begin{align}
&\frac{1}{4\pi i}\frac{1}{t\nu(0)\sqrt{\Omega}}\frac{1}{\beta} \sum_{\alpha k \sigma \omega_{n}}\left(\mathcal{ G}_{d_{\sigma}c_{\alpha k\sigma}^\dag}(i\omega_{n})-\mathcal{ G}_{c_{\alpha k\sigma}d_{\sigma}^\dag}(i\omega_{n})\right)=-\frac{1}{2\pi et^{2}\nu(0)}\sum_{\alpha}\alpha I_{\alpha} =0,
\end{align}
since in the steady state we must satisfy $I_{1}=I_{-1}$. Therefore, this results in:
\begin{align}
n_{d}&=\frac{1}{4\pi i}\sum_{\alpha\omega_{n}\sigma}\frac{1}{\beta}\int_{-\infty}^{\infty}d\epsilon_{k}
\bigg(\left\langle\left\{d^{\dag}_{\sigma},\frac{e^{i\omega_{n}0^{{+}}}}{i\omega_{n}+\mathcal{ L}-\mathcal{ L}_{Y}}\frac{1}{\epsilon_k+\mathcal{ L}-i\eta}d_{\sigma}\right\}\right\rangle\notag\\&\qquad\qquad-\left\langle\left\{\frac{e^{i\omega_{n}0^{{+}}}}{i\omega_{n}-\mathcal{ L}+\mathcal{ L}_{Y}}\frac{1}{\epsilon_k-\mathcal{ L}+i\eta}d^{\dag}_{\sigma},d_{\sigma}\right\}\right\rangle\bigg)\notag\\
&=\frac{1}{4\pi i}\sum_{\alpha\omega_{n}\sigma}\frac{1}{\beta}\int_{-\infty}^{\infty}d\epsilon_{k}
\bigg(\left\langle\left\{\frac{e^{i\omega_{n}0^{{+}}}}{i\omega_{n}-\mathcal{ L}+\mathcal{ L}_{Y}}d^{\dag}_{\sigma},\frac{1}{\epsilon_k+\mathcal{ L}-i\eta}d_{\sigma}\right\}\right\rangle\notag\\
&\qquad\qquad-\left\langle\left\{\frac{e^{i\omega_{n}0^{{+}}}}{i\omega_{n}-\mathcal{ L}+i\eta}d^{\dag}_{\sigma},\frac{1}{\epsilon_k+\mathcal{ L}-\mathcal{ L}_{Y}}d_{\sigma}\right\}\right\rangle\bigg).
\label{eq:intermediate}
\end{align}

In the next step, we will make use of the identity
\begin{align}\label{new-ident}
&\int_{-\infty}^{\infty} d\epsilon \frac{1}{{\epsilon}\pm\mathcal{ L}\mp i\eta}\mathcal{O}=\pm i\pi \mathcal{O},
\end{align}
and we briefly outline its proof. It is convenient to use the spectral representation for this purpose and thus enumerate the set of eigenstates of $H$ by $\{|n\rangle\}$. With this, one obtains
\begin{align}
&\int_{-\infty}^{\infty}d\epsilon\, \langle m\vert\frac{1}{{\epsilon}\pm\mathcal{ L}\mp i\eta}\mathcal{O}\vert n\rangle=\int_{-\infty}^{\infty}d\epsilon\, \frac{1}{{\epsilon}\pm(\epsilon_{m}-\epsilon_{n})\mp i\eta}\langle m\vert\mathcal{O}\vert n\rangle d\epsilon.
\end{align}
After shifting the integration variable $\epsilon \rightarrow \epsilon \pm(\epsilon_{m}-\epsilon_{n})$, we can further simplify the above expression to yield
\begin{align}
&\int_{-\infty}^{\infty}d\epsilon \, \frac{1}{{\epsilon}\mp i\eta}\langle m\vert\mathcal{O}\vert n\rangle=\left(\text{P.V.}\left[\int_{-\infty}^{\infty}\frac{d\epsilon}{\epsilon}\right]\pm i\pi\right)\langle m\vert\mathcal{O}\vert n\rangle.
\end{align}
Noting that the principal value integral vanishes concludes the proof of Eq.\ \eqref{new-ident}. With this identity in place, Eq.\ \eqref{eq:intermediate} can be brought into its final form
\begin{align}
n_{d}&=\frac{1}{4 }\sum_{\alpha\omega_{n}\sigma}\frac{1}{\beta}\bigg[\left\langle\left\{d^{\dag}_{\sigma},\frac{e^{i\omega_{n}0^{{+}}}}{i\omega_{n}+\mathcal{ L}-\mathcal{ L}_{Y}}d_{\sigma}\right\}\right\rangle+\left\langle\left\{\frac{e^{i\omega_{n}0^{{+}}}}{i\omega_{n}-\mathcal{ L}+\mathcal{ L}_{Y}}d^{\dag}_{\sigma},d_{\sigma}\right\}\right\rangle\bigg]\notag\\
&=\frac{1}{\beta}\sum_{\omega_{n}\sigma}\mathcal{ G}_{d_{\sigma}d_{\sigma}^\dag}(i\omega_n)=\sum_{\sigma}\mathcal{ G}_{d_{\sigma}{d}_{\sigma}^\dag}(\tau=0).
\end{align}
This demonstrates that the Schwinger-Keldysh approach and the effective equilibrium formulation reproduce the same expression for the occupancy on the quantum dot.

\subsection{Derivation of $n_{d}$ using the Keldysh approach}
\label{app:3} 
The basic equation which will constitute the starting point of our proof is given by Eq. (15) in Ref.\ \cite{Meir_Wingreen}. The current in lead $\alpha=\pm1$ for spin projection $\sigma$ is given by
\begin{align}
I_{\al
\sigma}(t)=&-\alpha\frac{2e}{\hbar}\int_{-t_{0}}^{t}dt_{1}\int\frac{d\ep}{2\pi}\text{Im}\bigg\{e^{-i\ep(t_{1}-t)}\Gamma^{\al}(t,t_{1})\notag\\
&\qquad\quad\times\left[G^{<}_{\sigma\sigma}(t,t_{1})+f_{\al}(\ep)G^{r}_{\sigma\sigma}(t,t_{1})\right]\bigg\},
\end{align}
where we have restricted ourselves to a single level on the dot having a spin index $\sigma$.
The occupation number($n_{d_{\sigma}}$) of the dot in the state $\sigma$ obeys the following differential equation:
\begin{align}
\frac{dn_{d_{\sigma}}(t)}{dt}=\frac{1}{-e}\sum_{\al} \al I_{\al\sigma}.
\label{eq:number}
\end{align}
In steady state ({\it i.e.}, t=0) we get 
\begin{align}
\frac{dn_{d_{\sigma}}(t)}{dt}\bigg|_{t=0}=0.
\end{align}
This implies 
\begin{align}
\sum_{\alpha}\int_{-\infty}^{0}dt_{1}\int\frac{d\ep}{2\pi}&\text{Im}\bigg\{e^{-i\ep t_{1}}\Gamma^{\al}(0,t_{1})\left[G^{<}_{\sigma\sigma}(0,t_{1})+f_{\al}(\ep)G^{r}_{\sigma\sigma}(0,t_{1})\right]\bigg\}=0.
\label{eq:derivative}
\end{align}
Since in steady state there exists time translational invariance, we have $G^{<}_{\sigma\sigma}(0,t_{1})=G^{<}_{\sigma\sigma}(0-t_{1})$ and $G^{r}_{\sigma\sigma}(0,t_{1})=G^{r}_{\sigma\sigma}(0-t_{1})$. Further we can set $\eta\rightarrow 0$ and $t_{0}\rightarrow -\infty$, in the sense defined by the open system limit, right from the beginning. This in turn implies that $\Gamma^{\al}(0,t_{1})=\Gamma^{\al}=\pi\nu(0)t^{2}=\Gamma/2$. Let us focus on the first term in the expression above
\begin{align}
\mathcal{ A}=&\sum_{\alpha}\int_{-\infty}^{0}dt_{1}\int\frac{d\ep}{2\pi}\text{Im}\bigg\{e^{-i\ep t_{1}}\Gamma^{\al}G^{<}_{\sigma\sigma}(-t_{1})\bigg\}\notag\\
&=-\frac{i}{2}\sum_{\alpha}\Gamma^{\al}\int_{-\infty}^{0}\int\frac{d\ep}{2\pi}\bigg\{e^{-i\ep t_{1}}G^{<}_{\sigma\sigma}(-t_{1})-e^{i\ep t_{1}}\left[G^{<}_{\sigma\sigma}(-t_{1})\right]^{\ast}\bigg\}\notag\\
\intertext{Using the fact $\left[G^{<}_{\sigma\sigma}(-t_{1})\right]^{\ast}=-G^{<}_{\sigma\sigma}(t_{1})$}
\mathcal{ A}=&-\frac{i}{2}\sum_{\alpha}\Gamma^{\alpha}\int\frac{d\ep}{2\pi}G^{<}_{\sigma\sigma}(\ep)=-\frac{i}{2}\Gamma G^{<}_{\sigma\sigma}(0)=\frac{1}{2}\Gamma \left\langle d^{\dag}(0)d(0)\right\rangle=\frac{1}{2}\Gamma n_{d_{\sigma}}.
\end{align}
Similarly for the second  term we get
\begin{align}
\mathcal{ B}=&\sum_{\alpha}\int_{-\infty}^{0}dt_{1}\int\frac{d\ep}{2\pi}\text{Im}\bigg\{e^{-i\ep t_{1}}f_{\al}(\ep)\Gamma^{\al}G^{r}_{\sigma\sigma}(-t_{1})\bigg\}\notag\\
&=\frac{\Gamma}{2}\sum_{\alpha}\int\frac{d\ep}{2\pi}f_{\al}(\ep)\text{Im}\bigg\{\int_{-\infty}^{0}dt_{1}e^{i\ep t_{1}}G^{r}_{\sigma\sigma}(-t_{1})\bigg\}\notag\\
&=\frac{\Gamma}{2}\sum_{\alpha}\int\frac{d\ep}{2\pi}f_{\al}(\ep)\text{Im}\bigg\{G^{r}_{\sigma\sigma}(\ep)\bigg\}=\Gamma\int\frac{d\ep}{2}\underbrace{\sum_{\alpha}\frac{f_{\al}(\ep)}{2}}_{f^{\text{eff}}(\ep,\Phi)}\underbrace{\frac{1}{\pi}\text{Im}\bigg\{G^{r}_{\sigma\sigma}(\ep)\bigg\}}_{-A_{d_{\sigma}}(\ep)}\notag\\
&=-\frac{1}{2}\Gamma\int d\ep f^{\text{eff}}(\ep,\Phi) A_{d_{\sigma}}(\ep).
\end{align}
 Combining the two terms and putting in Eq.\ (\ref{eq:derivative}) we get
\begin{align}
n_{d_{\sigma}}=\int d\ep f^{\text{eff}}(\ep,\Phi) A_{d_{\sigma}}(\ep),
\end{align}
which implies that
\begin{align}
n_{d}=\sum_{\sigma}n_{d_{\sigma}}&=\int d\ep f^{\text{eff}}(\ep,\Phi) \sum_{\sigma}A_{d_{\sigma}}(\ep)\notag\\
&=\int d\ep f^{\text{eff}}(\ep,\Phi) A_{d}(\ep).
\end{align}
This is identical to the expression obtained via the spectral representation in Appendix \ref{app:2}. 

 It is important to mention at this point that the expression for the local charge in terms of the spectral function in \eqr{eq:occupation} depends crucially on the fact that the system is in steady state.  Unlike its equilibrium counterpart, the equation $\sigma_{<}(\omega)=i[\sigma_{r}(\omega)-\sigma_{a}(\omega)]f^{\text{eff}(\omega)}$ does not hold in general \cite{Wilkins,Oguri_review}.  Here $\sigma_{<}(\omega)$, $\sigma_{r}(\omega)$ and $\sigma_{a}(\omega)$ denote the self energy contribution due to interactions to $G^{<}(\omega)$, $G^{r}(\omega)$ and $G^{a}(\omega)$ respectively. Using Dyson's equations in Appendix\ \ref{app:3}, it is simple to show that the weaker relation
\begin{align}
 \int_{-\infty}^{\infty}d\epsilon \left[\frac{\sigma_{<}(\epsilon)-i[\sigma_{r}(\epsilon)-\sigma_{a}(\epsilon)]f^{\text{eff}}}{(\epsilon-\epsilon_{d}-\rre[\sigma_{r}(\epsilon)])^{2}+(\Gamma-\iim[\sigma_{r}(\epsilon)])^{2}}\right]=0,
\end{align}
holds in steady state and leads to the equilbrium-like form of $n_{d}$ given by \eqr{eq:occupation}.

\section{Second-order perturbation theory details}
\label{app:second_order}
In this appendix we will outline the steps involved in the evaluation of the retarded electron Green's function to second-order in $H_{\text{int}}$ of Section 5.3. First we evaluate the relevant $D^{(k)}_{\sigma}(\omega)$'s involved in this computation. In the following discussion we will use the abbreviated notation which was introduced after Eq.\ \eqref{eq:abb}, {\it i.e.}, $l\equiv(\al_{l}k_{l})$, $\bar{l}\equiv(\al_{\bar{l}}k_{\bar{l}})$, $g_{l}\equiv g(\ep_{k_{l}})$ and $g_{\bar{l}}\equiv g(\ep_{k_{\bar{l}}})$. 

\begin{align}
D^{(0)}_{\sigma}(\omega)=\frac{t}{\sqrt{\Omega}}\sum_{1}\frac{g_{1}}{\omega-\ep_{1}+i\eta}\psi^{(0)}_{1\sigma},
\end{align}
follows directly from the definition. Using Eq. \eqref{eq:recursion} we get
\begin{align}
D^{(1)}_{\sigma}(\omega)=-\frac{1}{\omega+\mathcal{ L}'+i\eta}\mathcal{ L_{\text I}}D^{(0)}_{\sigma}.
\label{eq:B2}
\end{align}
To compute this quantity we first evaluate the action of the superoperator $\mathcal{ L}_{\text I}$ on 
$\psi_{\bar{1}\sigma}$.
\begin{align}
\mathcal{ L}_{\text I}\psi_{\bar{1}\sigma}
=-U&g^{\ast}_{\bar{1}}\left(\frac{t^{2}}{\Omega}\sum_{123}g^{\ast}_{1}g_{2}g_{3}\psi^{(0)\dag}_{1-\sigma}\psi^{(0)}_{2-\sigma}\psi^{(0)}_{3\sigma}-\frac{1}{2}\sum_{1}g_{1}\psi^{(0)\dag}_{1\sigma}\right).
\end{align}
Using this expression in Eq. \eqref{eq:B2} we obtain
\begin{align}
D^{(1)}_{\sigma}(\omega)&=U\frac{t}{\sqrt{\Omega}}\underbrace{\left(\frac{t^{2}}{\Omega}\right)\sum_{\bar{1}}\frac{\left|g_{\bar{1}}\right|^{2}}{\omega-\epsilon_{\bar{1}}+i\eta}}_{g(\omega)}\bigg(-\frac{1}{2}\sum_{1}\frac{g_{1}}{\omega-\ep_{1}+i\eta}\psi^{(0)\dag}_{1\sigma}\notag\\
&\qquad\qquad+\frac{t^{2}}{\Omega}\sum_{123}\frac{g^{\ast}_{1}g_{2}g_{3}}{\omega+\ep_{1}-\ep_{2}-\ep_{3}+i\eta}\psi^{(0)\dag}_{1-\sigma}\psi^{(0)}_{2-\sigma}\psi^{(0)}_{3\sigma}\bigg)\notag\\
&=U\frac{t}{\sqrt{\Omega}}g(\omega)\bigg(\frac{t^{2}}{\Omega}\sum_{123}\frac{g^{\ast}_{1}g_{2}g_{3}}{\omega+\ep_{1}-\ep_{2}-\ep_{3}+i\eta}\psi^{(0)\dag}_{1-\sigma}\psi^{(0)}_{2-\sigma}\psi^{(0)}_{3\sigma}\notag\\
&\qquad\qquad\qquad\qquad\qquad\qquad-\frac{1}{2}\sum_{1}\frac{g_{1}}{\omega-\ep_{1}+i\eta}\psi^{(0)\dag}_{1\sigma}\bigg).
\label{eq:d1}
\end{align}
The anticommutator of interest is then readily computed to give
\begin{align}
\left\{D^{(1)}_{\sigma}(\omega),d^{\dag}_{\sigma}\right\}=Ug(\omega)\left[-\frac{1}{2}g(\omega)+\frac{t^{2}}{\Omega}\sum_{12}g^{\ast}_{1}g_{2}g(\omega+\ep_{1}-\ep_{2})\psi^{(0)\dag}_{1-\sigma}\psi^{(0)}_{2-\sigma}\right].
\label{eq:A}
\end{align}

Similarly $D^{(2)}_{\sigma}(\omega)$ follows from the use of Eq.\ \eqref{eq:recursion} in Eq.\ \eqref{eq:d1} above
\begin{align}
D^{(2)}_{\sigma}(\omega)=-\frac{1}{\omega+\mathcal{ L}'+i\eta}\mathcal{ L_{\text I}}D^{(1)}_{\sigma}.
\end{align}

\noindent We begin by computing 
\begin{align}
\mathcal{ L_{\text I}}D^{(1)}_{\sigma}&=\frac{U}{2}\bigg[-\frac{t^{2}}{\Omega}\sum_{12\bar{\sigma}}g^{\ast}_{1}g_{2}\psi^{(0)\dag}_{1\bar{\sigma}}\psi^{(0)}_{2\bar{\sigma}}+\left(\frac{t^{2}}{\Omega}\right)^{2}\sum_{1234\bar{\sigma}}g^{\ast}_{1}g^{\ast}_{2}g_{3}g_{4}\psi^{(0)\dag}_{1\bar{\sigma}}\psi^{(0)\dag}_{2-\bar{\sigma}}\psi^{(0)}_{3-\bar{\sigma}}\psi^{(0)}_{4\bar{\sigma}},D^{(1)}_{\sigma}\bigg]\notag\\
&=-\frac{U}{2}\bigg[\frac{t^{2}}{\Omega}\sum_{12\bar{\sigma}}g^{\ast}_{1}g_{2}\psi^{(0)\dag}_{1\bar{\sigma}}\psi^{(0)}_{2\bar{\sigma}},D^{(1)}_{\sigma}\bigg]\notag\\
&\qquad\qquad\qquad+\frac{U}{2}\bigg[\left(\frac{t^{2}}{\Omega}\right)^{2}\sum_{1234\bar{\sigma}}g^{\ast}_{1}g^{\ast}_{2}g_{3}g_{4}\psi^{(0)\dag}_{1\bar{\sigma}}\psi^{(0)\dag}_{2-\bar{\sigma}}\psi^{(0)}_{3-\bar{\sigma}}\psi^{(0)}_{4\bar{\sigma}},D^{(1)}_{\sigma}\bigg].
\label{eq:temp}
\end{align}

It is convenient to divide the action of the Liouvillian above into two parts by defining 
\begin{align}
&\mathcal{ A}=-\frac{U}{2}\bigg[\frac{t^{2}}{\Omega}\sum_{12\bar{\sigma}}g^{\ast}_{1}g_{2}\psi^{(0)\dag}_{1\bar{\sigma}}\psi^{(0)}_{2\bar{\sigma}},D^{(1)}_{\sigma}\bigg]\notag\\
&\mathcal{ B}=\frac{U}{2}\bigg[\left(\frac{t^{2}}{\Omega}\right)^{2}\sum_{1234\bar{\sigma}}g^{\ast}_{1}g^{\ast}_{2}g_{3}g_{4}\psi^{(0)\dag}_{1\bar{\sigma}}\psi^{(0)\dag}_{2-\bar{\sigma}}\psi^{(0)}_{3-\bar{\sigma}}\psi^{(0)}_{4\bar{\sigma}},D^{(1)}_{\sigma}\bigg].
\end{align}

\noindent We introduce the notation
\begin{align}
&D^{(2)}_{\sigma\mathcal{ (A)}}(\omega)=-\frac{1}{\omega+\mathcal{ L}'+i\eta}\mathcal{ L_{\text I}}\mathcal{ A} \notag\\
&D^{(2)}_{\sigma\mathcal{ (B)}}(\omega)=-\frac{1}{\omega+\mathcal{ L}'+i\eta}\mathcal{ L_{\text I}}\mathcal{ B},
\end{align}
to separate the contributions to $D^{(2)}_{\sigma}(\omega)$ which follow from the parts $\mathcal{ A}$ and $\mathcal{ B}$.

\noindent The evaluation of the Green's function is likewise separated into two parts. First, we compute the contribution of the part $\mathcal{ A}$  
\begin{align}
&G^{\text{ret}(2)}_{d_{\sigma}d^{\dag}_{\sigma}(\mathcal{ A})}(\omega)=\left\langle\left\{D^{(2)}_{\sigma(\mathcal{ A})}(\omega),d^{\dag}_{\sigma}\right\}\right\rangle^{(0)}\notag\\
&\quad=\frac{U^{2}}{2}\left[g(\omega)\right]^{2}\bigg[\frac{\Gamma}{\pi}\int{d\ep_{1}}\left|g_{1}\right|^{2}\left(\frac{1}{\omega-\ep_{1}+2i\Gamma}-\frac{1}{\omega+\ep_{1}+2i\Gamma}\right)f^{\text{eff}}_{1}\bigg].
\label{eq:bitA}
\end{align}

\noindent Similarly, we compute the the contribution of part $\mathcal{ B}$ to the retarded electron Green's function of the dot $G^{\text{ret}(2)}_{d_{\sigma}d^{\dag}_{\sigma}}(\omega)$
\begin{align}
&G^{\text{ret}(2)}_{d_{\sigma}d^{\dag}_{\sigma}(\mathcal{ A})}(\omega)=-U^{2}\left[g(\omega)\right]^{2}\bigg[3\left(\frac{\Gamma}{\pi}\right)^{2}\int{d\ep_{1}}\int{d\ep_{2}}\left|g_{1}\right|^{2}\left|g_{2}\right|^{2}g(\omega+\ep_{1}-\ep_{2})f_{1}^{\text{eff}}f_{2}^{\text{eff}}\notag\\
&\qquad\qquad\qquad\qquad-\frac{1}{2}\frac{1}{\omega+3i\Gamma}-\frac{\Gamma}{\pi}\int{d\ep_{1}}\left|g_{1}\right|^{2}\left(\frac{1}{\omega+\ep_{1}+2i\Gamma}\right)f_{1}^{\text{eff}}\bigg].
\label{eq:bitB}
\end{align}
Combining Eqs.\ \eqref{eq:bitA} and \eqref{eq:bitB} we finally obtain the second order correction to the retarded electron Green's function of the dot
\begin{align}
G^{\text{ret}(2)}_{d_{\sigma}d^{\dag}_{\sigma}}(\omega)&=\ G^{\text{ret}(2)}_{d_{\sigma}d^{\dag}_{\sigma}(\mathcal{ A})}(\omega)+G^{\text{ret}(2)}_{d_{\sigma}d^{\dag}_{\sigma}(\mathcal{ B})}(\omega)\notag\\
&=U^{2}\left[g(\omega)\right]^{2}\bigg[\frac{1}{\omega+3i\Gamma}\notag\\
&\qquad-3\left(\frac{\Gamma}{\pi}\right)^{2}\int{d\ep_{1}}\int{d\ep_{2}}\left|g_{1}\right|^{2}\left|g_{2}\right|^{2}g(\omega+\ep_{1}-\ep_{2})f_{1}^{\text{eff}}f_{2}^{\text{eff}}\bigg].
\end{align}

\section{Deviations from the particle-hole symmetric point: Summary of results to $\mathcal{ O}(U^{2})$}
\label{app:deviation}

We study the effect a deviation from the particle-hole symmetric point for the Anderson impurity
\begin{align}
H&=\sum_{\sigma}\delta_{\sigma}d^{\dag}_{\sigma}d_{\sigma}+\frac{t}{\sqrt{\Omega}}\sum_{\alpha k\sigma}\left[d^{\dag}_{\sigma}c_{\al k\sigma}+c^{\dag}_{\al k\sigma}d_{\sigma}\right]\notag\\
&+\sum_{\alpha k\sigma}\ep_{k}c^{\dag}_{\alpha k\sigma}c_{\alpha k\sigma}+\frac{U}{2}\left(\widehat{n}_{d}-1\right)^2.
\end{align}
Here $\delta_{\sigma}$ measures the deviation from the particle-hole symmetric point. For a local magnetic field $H$ on the dot, we have
\begin{align}
\delta_{\sigma}=-\sigma\frac{H}{2}.
\end{align}
The retarded electron Green's function of the dot now bears an explicit $\sigma$ dependence, {\it i.e.}
\begin{align}
g_{d_{\sigma}}(\omega)=\frac{1}{\omega-\delta_{\sigma}+i\Gamma}.
\end{align}
We define the auxiliary function
\begin{align}
\tilde{g}_{d_{\sigma}}(\omega)=\frac{1}{\omega+\delta_{\sigma}+i\Gamma}.
\end{align}

Furthermore, the deviation of the occupancy of the dot level given by the spin projection $\sigma$ is denoted by
\begin{align}
\delta n_{d_{\sigma}}=\frac{\Gamma}{\pi}\int_{-\infty}^{\infty}d\ep\frac{1}{(\ep-\delta_{\sigma})^2+\Gamma^2}f^{\text{eff}}(\ep)-\frac{1}{2}.
\end{align} 

The first-order contribution to the complete self-energy $\bar{\Sigma}_{\sigma}^{\text{ret}(2)}$ (defined in Eq.\ \eqref{complete_sigma}) is frequency independent and is given by
\begin{align}
\bar{\Sigma}^{(1)}_{\sigma}=U\delta n_{d_{\sigma}}.
\end{align}

In The second-order contribution to the  {\it complete} self-energy consists of 3 parts. The contribution following from $\mathcal{ A}$ in Appendix \ref{app:second_order} gets modified to 
\begin{align}
\bar{\Sigma}^{(2)}_{\sigma(\mathcal{ A})}&=\frac{U^{2}}{2}\bigg[\frac{\Gamma}{\pi}\int d\ep_{1}\left|g_{1-\sigma}\right|^{2}f^{\text{eff}}_{1}\bigg(\frac{1}{\omega-\ep_{1}-\delta_{\sigma}+\delta_{-\sigma}+2i\Gamma}\notag\\
&\qquad-\frac{1}{\omega+\ep_{1}-\delta_{\sigma}-\delta_{-\sigma}+2i\Gamma}\bigg)-g_{d_{\sigma}}(\omega)\delta n_{d_{-\sigma}}\bigg],
\end{align}
whereas the contribution which follows from $\mathcal{ B}$ gets altered to

\begin{align}
\bar{\Sigma}^{(2)}_{\sigma(\mathcal{ B})}&=-U^{2}\bigg[\left(\frac{\Gamma}{\pi}\right)^2\int d\ep_{1}\int d\ep_{2}\left|g_{1\sigma}\right|^2\left|g_{2-\sigma}\right|^2 f^{(\text{eff})}_{1}f^{(\text {eff})}_{2}\bigg(\frac{1}{\omega-\ep_{2}-\delta_{\sigma}+\delta_{-\sigma}+2i\Gamma}\notag\\
&-\frac{1}{\omega+\ep_{2}-\delta_{\sigma}-\delta_{-\sigma}+2i\Gamma}
+g_{d_{-\sigma}}(\omega+\ep_{2}-\ep_{1})-\tilde{g}_{d_{-\sigma}}(\omega-\ep_{2}-\ep_{1})\bigg)\notag\\
&+\left(\frac{\Gamma}{\pi}\right)^2\int d\ep_{1}\int d\ep_{2}\left|g_{1-\sigma}\right|^2\left|g_{2-\sigma}\right|^2 f^{(\text{eff})}_{1}f^{(\text {eff})}_{2}g_{d_{\sigma}}(\omega+\ep_{1}-\ep_{2})-g_{d_{\sigma}}(\omega)\delta n_{d_{-\sigma}}\notag\\
&\times\left(\frac{\Gamma}{\pi}\int d\ep_{1}\left|g_{1-\sigma}\right|^2 f_{1}^{\text{eff}}\right)-\frac{\Gamma}{\pi}\int d\ep_{1}\left|g_{1-\sigma}\right|^2f_{1}^{\text{eff}}\bigg(\frac{1}{\omega+\ep_{1}-\delta_{\sigma}-\delta_{-\sigma}+2i\Gamma}\bigg)\bigg].
\end{align}
Finally we now get a second-order contribution from the term $\left\langle\left\{D_{\sigma}^{(1)}(\omega),d^{\dag}_{\sigma}\right\}\right\rangle^{(1)}$
\begin{align}
\bar{\Sigma}^{(2)}_{\sigma(\mathcal{ C})}&=\frac{U^{2}}{g_{d_{\sigma}}(\omega)}\bigg[\frac{\Gamma}{\pi}\int d\ep_{1}f^{(\text{eff})}_{1}\left\{\frac{\left|g_{1\sigma}\right|^2 g_{1-\sigma}}{\omega+\ep_{1}-\delta_{\sigma}-\delta_{-\sigma}+2i\Gamma}+\frac{\left|g_{1-\sigma}\right|^2 g^{\ast}_{1\sigma}}{\omega-\ep_{1}+2i\Gamma}\right\}\bigg]\delta n_{d_{\sigma}}.
\end{align}
The contribution $\bar{\Sigma}^{(2)}_{\sigma(\mathcal{ C})}$, corresponds to two concatenated Hartree-Fock diagrams, and is thus not contribute to the one-particle irreducible (1PI) diagrams, which define the self-energy $\Sigma^{\text{ret}}$ (see Eq.\ \eqref{Dyson1}).\\

\noindent In other words, for the case of a local magnetic field on the dot, the self-energy to second-order is modified to 
\begin{align}
\Sigma^{\ast}_{\sigma}=\bar{\Sigma}^{(1)}_{\sigma}+\bar{\Sigma}^{(2)}_{\sigma(\mathcal{ A})}+\bar{\Sigma}^{(2)}_{\sigma(\mathcal{ B})}.
\end{align}

\end{appendix}

\begin{figure}[t]
  \includegraphics[width=0.9\columnwidth]{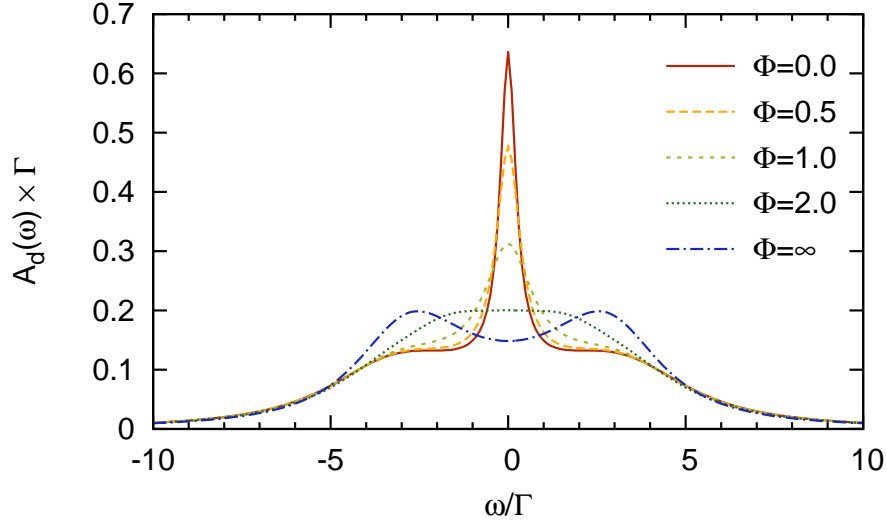}
	\caption{Spectral function (obtained using the Born approximation) for different values of the bias voltage for $U/(\pi\Gamma)=2$, where we set $\Gamma=1$. The limits $\Phi/\Gamma\rightarrow 0$ and $\Phi/\Gamma\rightarrow \infty$ agree with the results obtained via the Schwinger-Keldysh scheme \cite{Oguri_2002,Oguri_2001}.}
	\label{fig:SpectralFunction}
\end{figure}

\begin{figure}[h]
  \includegraphics[width=0.9\columnwidth]{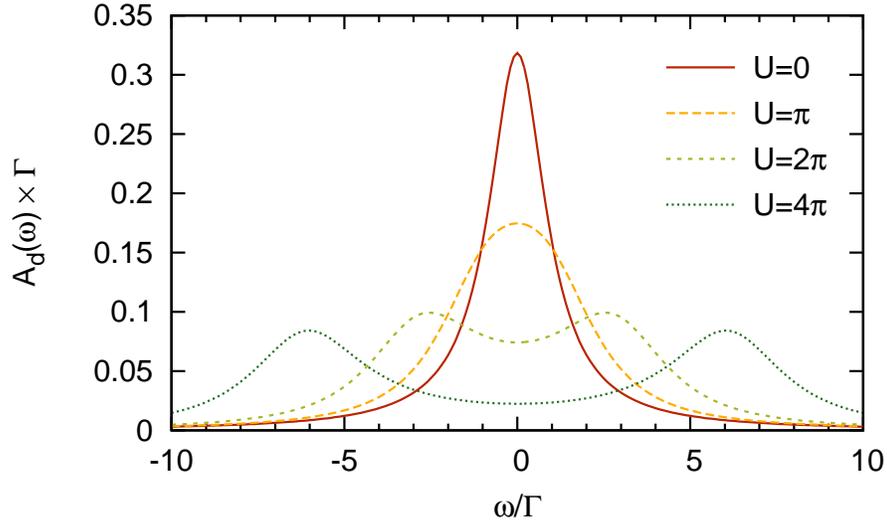}
	\caption{Behavior of the spectral function (obtained using the Born approximation) for different values of $U$, in the extreme limit $\Phi/\Gamma\rightarrow \infty$, where we set $\Gamma=1$. The results are identical to the results obtained using the Schwinger-Keldysh formalism \cite{Oguri_2002}.}
	\label{fig:SpectralFunction_1}
\end{figure}

\begin{figure}[h]
  \includegraphics[width=0.9\columnwidth]{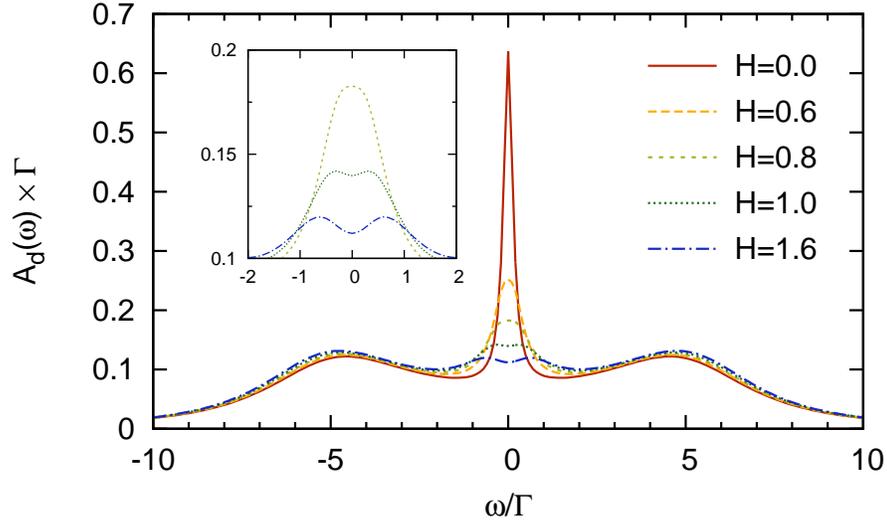}
	\caption{Variation of the spectral function with an applied magnetic field ($H$), obtained using the second-order self-energy. Here $\Gamma=1$, $\Phi=0$, $U=3\pi$ and $H$ is given in the units defined by $g\mu_{b}$, where $g$ is the g-factor and $\mu_{b}$ denotes the Bohr magneton. The inset shows the behavior of the Abriksov-Suhl resonance as a function of the magnetic field, for  $H=0.8$, $1.0$ and $1.6$.
	}
	\label{fig:magnetic_field}
\end{figure}

\begin{figure}[h]
  \includegraphics[width=0.9\columnwidth]{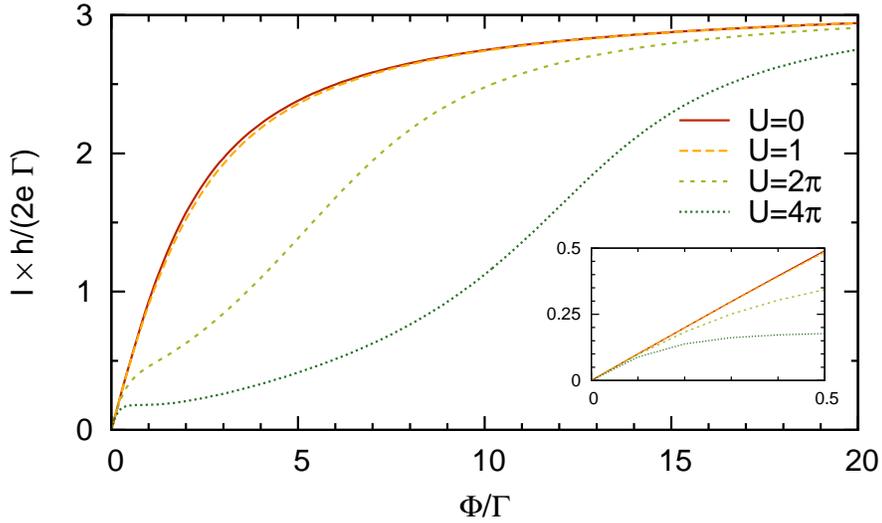}
	\caption{The current-voltage curves for the Anderson model for $U/\Gamma=0.0, 1.0, 2\pi$ and $4\pi$, where $\Gamma=1$. The inset shows the behavior of the curves for low bias. In the limit $\Phi\rightarrow 0$ the slope of the curves tend to 1, which corresponds to the value of the conductance quantum.}
	\label{fig:current_variation}
\end{figure}

\end{document}